\documentclass[12pt]{article}

\usepackage[english]{babel}

\usepackage[letterpaper,top=2cm,bottom=2cm,left=3cm,right=3cm,marginparwidth=1.75cm]{geometry}

\usepackage{amsmath,amssymb,bm}
\usepackage{amsthm}
\usepackage{graphicx}
\usepackage[colorlinks=true, allcolors=blue]{hyperref}
\usepackage{lscape}
\usepackage[shortlabels]{enumitem}
\usepackage{xcolor}
\usepackage{cleveref}
\usepackage{comment}
\usepackage{multicol}
\usepackage{forest}
\usepackage{tikz}
\usepackage{nicematrix}
\newcommand{\blue}[1]{\color{blue} #1 \color{black}}
\DeclareMathOperator{\im}{Im}

\forestset{
    roof/.style={edge path’={%
        (.parent first)--(!u.children)--(.parent last)--cycle
        }
    }
}

\newcommand{\A}{\mathtt{a}}
\newcommand{\C}{\mathtt{c}}
\newcommand{\G}{\mathtt{g}}
\newcommand{\T}{\mathtt{t}}
\renewcommand{\aa}{u^1}
\newcommand{\cc}{u^3}
\renewcommand{\gg}{u^2}
\renewcommand{\tt}{u^4}

\newcommand{\pia}{\pi_{A}}
\newcommand{\pic}{\pi_{C}}
\newcommand{\pig}{\pi_{G}}
\newcommand{\pit}{\pi_{T}}

\newcommand{\rank}{\mathrm{rank}}
\newcommand{\diag}{\mathrm{diag}}
\renewcommand{\flat}{\mathrm{Flat}}
\newcommand{\flatt}{\mathrm{Flat}_{12\mid 34}}
\newcommand{\la}{\lambda_{1}}
\newcommand{\lc}{\lambda_{3}}
\renewcommand{\lg}{\lambda_{2}}
\newcommand{\lt}{\lambda_{4}}
\newcommand{\RR}{\mathbb{R}}
\newcommand{\CC}{\mathbb{C}}

\newcommand{\pe}[2]{\langle #1, #2 \rangle}
\newcommand{\pes}[2]{\langle #1, #2 \rangle_{\pi}}
\renewcommand{\k}{\kappa}
\newcommand{\p}[1]{\pi_{\mathtt{#1}}} 
\newcommand{\bpt}[3]{\bar{p}_{\mathtt{#1}\mathtt{#2}\mathtt{#3}}} 

\newcommand{\MC}[1]{{\color{blue} \sf  Marta: [#1]}}
\newcommand{\RH}[1]{{\color{olive} \sf  Roser: [#1]}}

\newtheorem{thm}{Theorem}[section]
\newtheorem{exmp}[thm]{Example}
\newtheorem{lemma}[thm]{Lemma}
\newtheorem{prop}[thm]{Proposition}
\newtheorem{cor}[thm]{Corollary}

\newtheorem{remark}[thm]{Remark}
\newtheorem{question}[thm]{Question}
\theoremstyle{definition}
\newtheorem{definition}[thm]{Definition}

\title{A novel algebraic approach to time-reversible evolutionary models}
\author{Marta Casanellas, Roser Homs Pons, Angélica Torres}

\begin{document}
\maketitle

\begin{abstract}
In the last years algebraic tools have been proven to be useful in phylogenetic reconstruction and model selection by means of the study of phylogenetic invariants. However, up to now, the models studied from  an algebraic viewpoint are either too general or too restrictive (as group-based models with a uniform stationary distribution) to be used in practice.

In this paper we provide a new framework to work with time-reversible models, which are the most widely used by biologists.  In our approach we consider algebraic time-reversible models on phylogenetic trees (as defined by Allman and Rhodes) and introduce a new inner product to make all transition matrices of the process diagonalizable through the same orthogonal eigenbasis.  This framework generalizes the Fourier transform widely used to work with group-based models and recovers some of the well known results. As illustration, we exploit the combination of our technique with algebraic geometry tools to provide relevant phylogenetic invariants for trees evolving under the Tamura-Nei model of nucleotide substitution.  
\end{abstract}

\section{Introduction}\label{sec:intro}

Phylogenetics aims at recovering the evolutionary history of a given set of biological species from certain molecular information. This evolutionary process is represented on a phylogenetic tree or network whose leaves correspond to living species and whose interior nodes represent their common ancestors.  One of the most common ways of approaching phylogenetic reconstruction is by modeling the substitution of molecular units (usually nucleotides or amino acids) via a Markov process on a phylogenetic tree.

During the last twenty years, algebraic methods have been developed with the aim of helping biologists address phylogenetic reconstruction. The key is that Markov processes on phylogenetic trees parameterize algebraic varieties and tools from algebraic geometry turn out to be relevant as suggested by Felsenstein, Cavender, and Lake in the late eighties, see \cite{Cavender87, Lake1987}. They introduced the use of \emph{phylogenetic invariants} which are polynomial constraints satisfied by any distribution that arises as a hidden Markov process on a phylogenetic tree. These tools avoid parameter inference, which might be a tedious task, and incorporate the geometry of the algebraic varieties to detect the tree that best fits the given data, in a certain measure. Methods based on algebraic tools such as SVDquartets \cite{chifmankubatko2014} or Erik+2 \cite{fercas2016} have been implemented successfully in the phylogenetic software PAUP* \cite{paup}. These methods consider the most \emph{general Markov} model of nucleotide substitution (GM for short). Other models that have been studied by algebraists are $G$-equivariant models (see \cite{Draisma}, \cite{CFM}, \cite{CF11}), which are friendly models from a mathematical approach but only used by biologists in very special cases.

Markov processes on phylogenetic trees mostly used by biologists have the property of being \emph{stationary}. The GM model on a phylogenetic tree is not a stationary process and, among $G$-equivariant models, those that are stationary are too simple as their stationary distribution is uniform. Another property that is commonly assumed by biologists is \emph{time-reversibility}. Roughly speaking, a stationary Markov process is time-reversible if, at equilibrium, the rate at which transitions from state $i$ to state $j$ occur is the same as the rate at which transitions from $j$ to $i$ occur.  
Thus, there is a need to provide algebraic methods for time-reversible processes on phylogenetic trees, for any stationary distribution. This can be specially relevant in the case of amino acid substitution models, where the GM model is too large to be of biological utility.

The time-reversibility property has been studied from an algebraic point of view in \cite{AR_stationary}. Allman and Rhodes tailor time-reversibility from an algebraic approach and define the class of \emph{algebraic time-reversible} models (ATR briefly). This class contains all time-reversible models that biologists use in their everyday work such as GTR \cite{tavare86}, TN93 \cite{TN93_p}, or HKY85 \cite{HKY85_p}. In an ATR model on a tree, all transition matrices must commute. This is a natural requirement since it is satisfied by all time-homogeneous continuous-time processes, which are the most widely used in phylogenetic reconstruction.

We build upon this definition of ATR models and develop a new framework that simplifies the study of these models. This can be thought as a generalization of the well studied Hadamard or Fourier transform for group-based models exploited in a large list of publications:  \cite{Evans1993}, \cite{SSE93}, \cite{Hendy1994}, and \cite{Sturmfels2005} among others. First of all, if data has reached equilibrium, the stationary distribution $\pi$ can be inferred from the data and we can consider it as input data (this approach was already considered by the first author and M. Steel in the study of the Equal-Input model, \cite{CasSteel}). Then, for a fixed stationary distribution $\pi$ of an ATR model on a phylogenetic tree, we introduce a new inner product $\langle \,,\rangle_{\pi}$ and prove that all transition matrices diagonalize under an orthogonal eigenbasis with respect to $\langle \,,\rangle_{\pi}$. By fixing this orthogonal eigenbasis, we are able to do a change of coordinates that simplifies the parameterization of our model. For example, we are able to recover the celebrated result of Evans and Speed \cite{Evans1993}.  With these new coordinates we can provide phylogenetic invariants for these Markov processes on trees and describe the corresponding algebraic varieties. We illustrate these tools with a deep study of the TN93 model and give phylogenetic invariants that can be used for topology reconstruction or model selection. We focus this study on quartets (i.e. trees with four leaves) because they can be used as a building piece in phylogenetic reconstruction by means of Quartet-based methods, see \cite{stjohn2003} for instance.

The structure of the paper is as follows. In \Cref{sec:preliminaries} we introduce the preliminaries on Markov processes on phylogenetic trees. In \Cref{sec:ATR} we develop the framework that allows us to disentangle ATR models on trees: we introduce the inner product  $\langle \,,\rangle_{\pi}$, prove that ATR models on trees deal with transition matrices that simultaneously diagonalize in a basis that is orthogonal for this inner product, and define algebraic varieties associated to these models. In \Cref{sec:coordinates} we explore the change of coordinates to this eigenbasis and prove the main technique that permits the study of these models on phylogenetic trees from the study on smaller trees (Theorem \ref{thm:gluing}). In \Cref{sec:TN93} we delve into the study of the TN93 model: we give phylogenetic invariants for trees evolving under this model with any number of leaves and, for quartet trees, we specify a collection of phylogenetic invariants that (locally) cut out the algebraic variety associated to this model. In other words, we give a collection of constraints that suffice to describe distributions evolving under a quartet tree under the TN93 model, see Theorem \ref{lem:localCIquartet}. 
{All computations are available in the institutional CORA repository  \url{https://dataverse.csuc.cat/dataset.xhtml?persistentId=doi:10.34810/data1128}


\textbf{Acknowledgments.} We would like to thank Jes\'us Fern\'andez-S\'anchez for useful discussions on this topic that led to improvements of the paper. 

\section{Preliminaries}\label{sec:preliminaries}

In this section we give a brief introduction to Markov processes on phylogenetic trees and set up some notation needed throughout the paper. These concepts can be found in \cite[chapters 1 and 8]{SteelPhylogeny}.

Let $L=\{l_1,\dots,l_n\}$ be a finite set of cardinality $n$ (in our setting these elements represent biological entities, such as homologous genes of different species). A \emph{phylogenetic tree $T$ on $L$}   is a tree (connected acyclic graph) with leaf set $L$ (that is, the leaf nodes of the graph are in bijection with $L$). 
We use  $E(T), N(T), Int(T)$ to denote the set of edges, nodes and interior nodes of $T$, respectively. We say that $T$ is a \emph{rooted} phylogenetic tree if we specify an interior node $r$ of $T$ and direct all edges away from it. If $e=u \rightarrow v$ is an edge on a rooted tree,  we say that $u$ is the \emph{parent node} of $e$ (denoted as $p(e)$), and $v$ is the child node of $e$ (denoted as $c(e)$). The set of all phylogenetic trees on $L$ will be denoted as $\mathcal{T}_n$.


Molecular sequences can be thought as ordered sequences of a finite set of characters or states. We call $\Sigma$ this finite set of $\kappa$ states and assume that different positions on the sequence are independent and identically distributed, so that we only model the evolution on one site. For example, we use $\Sigma=\{A,G,C,T\}$ if we consider nucleotide sequences or $\kappa=20$ if we consider amino acid sequences. We denote the elements of  $\Sigma$ by $\{1,\dots, \kappa\}$ for convenience.

We recall how to describe a Markov process on a phylogenetic tree $T$ to model the evolution of molecular sequences along $T$.
At each node $v$ of a rooted phylogenetic tree $T$ we assign a random  variable $X_v$ taking values in $\Sigma$. 
We introduce a Markov process on $T$ by defining a parametric statistical model which assumes that each random variable is conditionally independent of its non-descendants given its parent variable \cite[\S 3]{Piotrbook}. 
The parameters of these models are the distribution $\pi^r$ at the root node $r$ and a Markov or transition matrix $M^e$ for each edge $e=u\rightarrow v \in E(T)$.   The entry $i,j$ of the Markov matrix $M^e$ stands for the conditional probability of observing state $j \in \Sigma$ at $X_v$ given the observation of state $i\in \Sigma$ at $X_u$. 

By definition, the entries of a Markov matrix are conditional probabilities. However, in this work we extend this term to allow for negative entries. That is, by a $\kappa \times \kappa$ Markov matrix we mean a real square matrix whose rows sum to one. We denote as $\mathcal{M}^1_{\RR}$ the set of $\kappa\times\kappa$ Markov matrices ($\kappa$ will be understood from the context and 1 denotes that the sum of rows is equal to one) and  $\mathcal{M}^1_{\CC}$ is defined analogously. 

A \emph{character $\bm{i}$ on} $N(T)$ is an assignment of states at the nodes of $T$, that is $\bm{i}=(i_v)_{v\in N(T)}$, $i_v\in \Sigma$. If all Markov matrices $M^e$ are non-negative, the probability of observing a character $\bm{i}$ at the nodes of $T$ is  \begin{equation}\label{eq:full}
p_{\bm{i}}^T=\pi^r_{i_r}\prod_{e \in E(T)} M^e_{i_{p(e)},i_{c(e)}}
\end{equation}
and this expression can be extended to matrices in $\mathcal{M}_{\RR}^1$ or in $\mathcal{M}_{\CC}^1$.
%

If $A$ is a subset of the set of nodes and $\bm{j}_A=(j_v)_{v \in A}$, $j_v \in \Sigma$, is a collection of states at the nodes of $A$, we say that a character  $\bm{i}=(i_v)$ on $N(T)$ \emph{extends} $\bm{j}_A$ if $i_v=j_v$ for all $v\in A$. The set of all characters on $N(T)$ that extend $\bm{j}_A$ is denoted as $ext(\bm{j}_A)$. One defines
\begin{equation}\label{eq:prob}
p^T_{\bm{j}_A}=\sum_{\bm{i} \in ext(\bm{j}_A)}p^T_{\bm{i}}   
\end{equation}
as the marginalization over $N(T)\setminus A.$
When $A=L(T)$, we denote $p^T_{\bm{i}_A}$ as $p^T_{i_1\dots i_n}$ and in this case expression \eqref{eq:prob} can be rewritten as
\begin{equation*}
p^T_{i_1\dots i_n}=\sum_{ \begin{array}{c}\scriptstyle v\in Int(T) \\
\scriptstyle i_v \in \Sigma\end{array}} \pi^r_{i_r}\prod_{e \in E(T)} M^e_{i_{p(e)},i_{c(e)}}.    
\end{equation*}

If $F$ is a field (either $\RR$ or $\CC$), we call $W$ the $F$-vector space of dimension $\kappa$ and call $e^1,\dots, e^{\kappa}$ its standard basis.
This induces a standard basis in the tensor product  $\otimes^l W$, $l\geq 1$: $e^{i_1}\otimes\dots\otimes e^{i_l}$, $i_j\in \Sigma$. We often identify a tensor in $\otimes^l W$ with the column vector formed by its coordinates in the standard basis. In this way, a joint distribution $(p_{1\dots 1},\dots,p_{\kappa \dots \kappa})$ can be identified with a $n-way$ tensor in $\otimes^n W.$

If we call $\mathcal{P}$ the set of free parameters of the Markov process, the \emph{(hidden) Markov process on} $T$ is the map
\begin{equation}\label{eq_mapphi}
\begin{array}{rcl}
   \mathcal{P} & \stackrel{\phi_T}{\longrightarrow} &\bigotimes^n W  \\
     \pi^r,\left( M^{e}\right)_{e \in E(T)} & \mapsto & p^T= \sum_{i_1,\dots,i_n} p^T_{i_1\dots i_n} e^{i_1}\otimes\ldots \otimes e^{i_n},
\end{array}    
\end{equation}
which assigns the joint distribution at the leaves of the tree to each set of Markov matrices and each $\pi^r$.  
If no further restrictions on the Markov matrices or on the distribution at the root are assumed, then this is called a \emph{general Markov} process on $T$. We omit the superscript $T$ in $p^T$ when it is understood from the context and even if we could distinguish whether $F=\RR$ or $F=\CC$, we talk about the same map $\phi_T$.

One of the main constructions for studying the general Markov model from an algebraic viewpoint is the flattening of a tensor. We recall the definition below. 

\begin{definition}\label{def:flat} Let $A|B$ be a bipartition of the set of leaves $L$. Assume that leaves are ordered so that $A=\{1,\dots,m\}$, $B=\{m+1,\dots,n\}$. If $p\in \otimes^n W$, the \emph{flattening of $p$ according to the bipartition} $A|B$ is the $\kappa^{|A|}\times \kappa^{|B|}$ matrix $\flat_{A|B}(p)$ whose $(i_1\dots i_m,i_{m+1}\dots i_n)$ entry is $p_{i_1\dots i_n}.$ For any other order of the set of leaves, $\flat_{A|B}(p)$ is defined analogously.
\end{definition}

The following theorem is one of the main tools in algebraic phylogenetics.

\begin{thm}[\cite{Allman2008}]\label{thm:ARflat}
Let $T\in \mathcal{T}_n$ and let $p\in \im(\phi_T)$. Then, if $A|B$ is a bipartition induced by removing an edge of $T$, $\rank \left(\flat_{A|B}(p)\right)$ is bounded above by $ \kappa$. On the contrary, if $A|B$ cannot be induced by removing any edge of $T$,  the rank of $\flat_{A|B}(p)$ is larger than $\kappa$ if the parameters that gave rise to $p$ were sufficiently general.     
\end{thm}

\section{Time-reversible evolutionary models}\label{sec:ATR}

Let $\Delta^{\kappa-1}$ be the standard simplex in $\RR^\kappa$, fix a distribution $\pi \in \Delta^{\kappa-1}$ with non-zero entries, and call $D_{\pi}$ the diagonal matrix $diag(\pi)$.  Any positive Markov matrix has a unique stationary distribution (a left-eigenvector of eigenvalue 1) and we say that a  $\kappa \times \kappa$ Markov matrix $M$ is $\pi$-\emph{stationary} if $\pi^t M=\pi^t$. 
%

A Markov matrix $M$ is \emph{$\pi$-time-reversible} if $D_{\pi}M=M^t D_{\pi}$, 
that is, $\pi_{i}{M}_{i,j}=\pi_{j}{M}_{j,i}$ for any $i,j$. Note that if $M$ is $\pi$-time-reversible, then $M$ is $\pi$-stationary. In terms of probabilities, the time-reversibility condition means that the probability of observing state $i$ at the parent node and $j$  at the child node of the process governed by $M$ is the same as observing $i$ at the child node and $j$ at the parent node. 
We introduce an inner product that gives another way of expressing time-reversibility.
\begin{definition}
The $\pi$-\emph{inner product} of $u,v\in W$ is
\[\langle u,v\rangle_{\pi} :=\sum_i \frac{1}{\pi_i} u_i {v_i}=u^t D_{\pi}^{-1}{v},\]
where $u_i$ and $v_i$, for $i=1,\ldots ,\k$, are the coordinates of $u,v$ in the standard basis.    
\end{definition}

Then, $M$ is $\pi$-time-reversible if and only if $M^t$ is a self-adjoint matrix with respect to this inner product (that is, $\langle M^tu,v\rangle_{\pi}=\langle u,M^tv\rangle_{\pi}$ for any $u,v$). In particular, thanks to the Spectral Theorem, all eigenvalues of $M$ are real and there exists a basis of eigenvectors of $M^t$ which is orthogonal with respect to $\langle \; , \, \rangle_{\pi}$. An orthogonal basis with respect to $\langle \; , \, \rangle_{\pi}$ will be called a \emph{$\pi$-orthogonal basis}.

\begin{remark}\rm
  Be aware that in \cite[\S 12]{LevinPeresWilmer2006} a similar inner product $\langle\; , \, \rangle '$ is defined but using $D_{\pi}$ instead of $D_{\pi}^{-1}$. We have that $M^t$ is self-adjoint with respect to $\langle \; , \, \rangle_{\pi}$ if and only if $M$ is self-adjoint with respect to $\langle\; , \, \rangle '$.  We defined the inner product this way because we are interested in eigenvectors of $M^t$ instead of $M$ (as Markov matrices act to the right of row vectors). 
\end{remark}

A Markov process on a phylogenetic tree is $\pi$-time-reversible 
if all its transition matrices are $\pi$-time-reversible and
$\pi$ is the distribution at the root.  In \cite{AR_stationary}, all transition matrices of the process are assumed to commute with each other to say that the process is \emph{algebraic time-reversible} (ATR briefly). This extra assumption is equivalent to saying that  all matrices simultaneously diagonalize (if they are diagonalizable), and it is an implicit assumption when one considers continuous time-reversible models that are homogeneous over time (that is, for any edge $e$, $M^e=exp(t_eQ)$ for a certain rate matrix $Q$). As these are the most widely used processes in phylogenetic software, in this paper we consider ATR processes on trees. 

If we have an ATR process with stationary distribution $\pi$, then there exists a $\pi$-orthogonal basis 
\[B=\{u^1,\dots,u^{\kappa}\}\]
which diagonalizes all transpose matrices $(M^e)^t$, $e \in E(T)$. As $\pi$ is a left-eigenvector with eigenvalue 1 for each $\pi$-time-reversible Markov matrix $M^e$, we can assume $u^1=\pi$. In particular, $\langle {u^1},{u^1}\rangle_{\pi}=1$ and $\langle{u^1},{e^i}\rangle_{\pi}=1$ for any $i=1,\dots,\kappa$.

\begin{definition}\label{def:evol}
Let $\pi\in\Delta^{\kappa -1}$ be a fixed distribution with non-zero entries and let $B=\{u^1=\pi, u^2,\dots,u^{\kappa}\}$ be a $\pi$-orthogonal basis in $\RR^{\kappa}.$  A  phylogenetic tree $T$ \emph{evolves under a} $B$-\emph{time-reversible model} if all Markov matrices $M^e$, $e \in E(T)$ on the Markov process on $T$ have $B$ as a left eigenbasis and $\pi^r=\pi$. 
\end{definition}

The following lemma guarantees that a $B$-time-reversible model on a phylogenetic tree is an ATR process. Before proving it, we introduce some notation. Throughout this paper we denote by $\textbf{1}$ the vector $\sum_{j=1}^{\kappa}e^j$ and  let $A$ be the change of basis matrix from $B$ to the standard basis $e^1,\dots, e^{\kappa}$, that is, $A=\begin{pmatrix}
u^1 &\dots &u^{\kappa}
\end{pmatrix}.$ As $B$ is $\pi$-orthogonal we have
    \begin{equation}\label{eq_A}
        A^t D_{\pi}^{-1} A= S,
    \end{equation}
    where $S$ is the diagonal matrix  $\diag(\langle u^1, u^1 \rangle_{\pi},\dots, \langle u^{\kappa}, u^{\kappa}\rangle_{\pi}).$

\begin{lemma}\label{lem:rowSum}
Let $\pi\in\Delta^{\kappa -1}$ be a distribution with positive entries, $B=\{u^1=\pi,\dots,u^{\kappa}\}$ a $\pi$-orthogonal basis in $\RR^{\kappa}$, and $M$ a $\kappa\times \kappa$ matrix for which $B$ is a left eigenbasis.  

Then $D_{\pi}M=M^tD_{\pi}$ and $M$ has constant row sum; moreover, the first eigenvalue $\lambda_1$ is equal to one if and only if $M$ is a $\pi$-time-reversible Markov matrix (and in this case, $\pi$ is a stationary distribution for $M$).
\end{lemma}

\begin{proof}   
If $B$ is a left eigenbasis for $M$, we have $M= A^{-t}\Lambda A^t$ for some diagonal matrix $\Lambda$. Thus, $D_{\pi}M=D_{\pi}A^{-t}\Lambda A^t$, which equals $A S^{-1} \Lambda A^t$ by \eqref{eq_A}. As $S^{-1}$ and $\Lambda$ commute, applying \eqref{eq_A} again we have $D_{\pi}M=A \Lambda A^{-1}D_{\pi}$, which is $M^t D_{\pi}$ as we wanted to prove.
    
Note that as $\pe{u^1}{u^i}_\pi=0$ for any $i\neq 1$,  $u^i$ has sum of coordinates equal to 0. Thus,  the first column of $A$ adds to 1 and the other columns add to 0. In particular, if $\Lambda=\operatorname{diag}( \lambda_1,\dots,\lambda_{\kappa})$, we have 
    \[M\textbf{1}=A^{-t}\Lambda A^t \textbf{1} =A^{-t}\Lambda e^1=\lambda_ 1 A^{-t}e^1=\lambda_1 D_{\pi}^{-1}Ae^1= \lambda_1 D_{\pi}^{-1} u^1=\lambda_1\textbf{1}.\]
     Thus, $M$ has constant row sum.  
   Requiring sum of rows equal 1 on a square matrix is equivalent to saying that $\textbf{1}$ is an  eigenvector of eigenvalue 1, so the last claim also follows easily.
\end{proof}

\begin{exmp}\label{ex:TN93}\rm(Tamura-Nei model, TN93)
Tamura and Nei presented in \cite{TN93_p} a continuous-time model based on the observed changes in human mithocondrial DNA. They proposed an arbitrary stationary distribution $\pi$, and observed that probabilities of transitions (changes within purines or within pyrimidines) and transversions (changes between purines and pyrimidines)  depend on the frequencies of the obtained nucleotide and on a single parameter for transversions and two for transitions. This is a time-reversible model 
whose transition matrices have the form 

\begin{equation}
M=\begin{pmatrix}
*_{\mathtt{1}} &\pi_{2}c & \pi_{3}b & \pi_{4}b\\
\pi_{1}c &*_{2} & \pi_{3}b & \pi_{4}b\\
\pi_{1}b &\pi_{2}b & *_{3} & \pi_{4}d\\
\pi_{1}b & \pi_{2}b &\pi_{3}d &*_{4}
\end{pmatrix},
\end{equation}
where $*_i$ is chosen so that each row sums to 1. Here we identified $\Sigma$ with the set of nucleotides $\{A,G,C,T\}$, in this order.
The matrix $M^t$ has the following $\pi$-orthogonal basis $B$ of eigenvectors: 
\[B=\left\{u^1=\begin{pmatrix}{\pi_1}\\ {\pi_2}\\ {\pi_3}\\ {\pi_4}\end{pmatrix}, \,
u^2 =\begin{pmatrix} \pi_1 \pi_{34}\\  \pi_2\pi_{34}\\ -\pi_3 \pi_{12}\\ -\pi_4 \pi_{12}
\end{pmatrix},\,
u^3=\frac{1}{\pi_{34}}\begin{pmatrix}0\\  0 \\ {\pi_3}\pi_4\\-\pi_3\pi_4\end{pmatrix}, \,
u^4=\frac{1}{\pi_{12}}\begin{pmatrix} {\pi_1}\pi_2 \\-\pi_1\pi_2\\ 0 \\0\end{pmatrix}\right\},\]
where $\pi_{12}=\pi_1+\pi_2$ and $\pi_{34}=\pi_3+\pi_4$. If the columns of $A$ are the vectors of $B$, then 
\begin{equation}\label{eq:Mlambdas}
M^t=A\begin{pmatrix}\la &0&0&0\\
0&\lc&0&0\\
0&0&\lg&0\\
0&0&0&\lt
\end{pmatrix} A^{-1},
\end{equation}
where
$\lambda_{1}=1$,  $\lambda_{2}=\lambda_{1}-b$, $\lambda_{3}=\lambda_{1}-\pi_{12}b-\pi_{34}d$, $\lambda_{4}=\lambda_{1}-\pi_{34}b-\pi_{12}c$
are the eigenvalues of $M$. 
%
The entries of $M$ can be written in terms of the eigenvalues using that 
\begin{eqnarray*}
b=\lambda_1-\lambda_2, \quad 
c=\frac{\pi_{12}\lambda_1+{\pi_{34}}\lambda_2 -{\lambda_4}}{\pi_{12}}, \quad d=\frac{\pi_{34}\lambda_1+{\pi_{12}}\lambda_2 -{\lambda_3}}{\pi_{34}}, \\
*_1=\frac{\pi_1}{\pi_{12}}(\pi_{12}\lambda_1+{\pi_{34}}\lambda_2+\frac{\pi_{2}}{\pi_1}\lambda_4), \quad
*_2=\frac{\pi_2}{\pi_{12}}(\pi_{12}\lambda_1+{\pi_{34}}\lambda_2+\frac{\pi_{1}}{\pi_2}\lambda_4), \\
*_3=\frac{\pi_3}{\pi_{34}}(\pi_{34}\lambda_1+{\pi_{12}}\lambda_2+\frac{\pi_{4}}{\pi_2}\lambda_2), \quad
*_4=\frac{\pi_4}{\pi_{34}}(\pi_{34}\lambda_1+{\pi_{12}}\lambda_2+\frac{\pi_{3}}{\pi_4}\lambda_3).
\end{eqnarray*}


A matrix $M$ satisfying  \eqref{eq:Mlambdas} is an \emph{algebraic} TN93 matrix (and we do not necessarily assume $\lambda_1=1$). If we impose $c=d$ then we have an HKY85 matrix \cite{HKY85_p}, 
and if we impose $b=c=d$, we obtain the Equal-Input model (EI)\cite{Felsenstein2005_p,CasSteel}. If we adopt the extra assumption that the stationary distribution is uniform, $\pi=(1/4,1/4,1/4,1/4)$, we recover the RY3.3c model of \cite{Woodhams}.
Hence, TN93 is a $B$-time-reversible model and so are its submodels HKY85, EI and RY3.3c. 

Note that 
\[\pes{\aa}{\aa} =1,\quad \pes{\gg}{\gg}=\pi_{12}\pi_{34},\quad \pes{\cc}{\cc}=\frac{\pi_3\pi_4}{\pi_{34}},\quad \pes{\tt}{\tt}=\frac{\pi_1\pi_2}{\pi_{12}}\, .\] The following table gives the $\pi$-inner product among the vectors in the standard basis and  the basis $B$:
\begin{equation}\label{eq:table_scalar}
\begin{array}{c|cccc}
   \pes{\cdot}{\cdot} & u^1 & u^2&u^3&u^4  \\
   \hline
    e^1 & 1& \pi_{34} & 0 &\pi_2/\pi_{12}\\
    e^2 & 1 & \pi_{34} & 0 & -\pi_{1}/\pi_{12}\\
    e^3 & 1 & -\pi_{12}&\pi_4/\pi_{34} &0\\
    e^4 & 1 & -\pi_{12}&-\pi_3/\pi_{34} &0\\
\end{array}    .
\end{equation}
For any $B$-time-reversible model, as the standard basis is also a $\pi$-orthogonal basis, we have that $\langle u^i, e^j\rangle_{\pi}$ is the $j$-th coordinate of $u^i$ (in the standard basis) divided by $\pi_j$ (because $\langle e^j,e ^j\rangle_{\pi}=1/{\pi_j}$).
\end{exmp}

\begin{remark}\rm Note that there are $B$-time-reversible models that are not \emph{multiplicatively closed}. This important property has been argued to be needed for consistency of phylogenetic inference, see \cite{Sumner_gtr}. For instance, regarding the models in the previous example, while TN93 is multiplicatively closed, its submodel HKY85 is not (the product of two HKY85 matrices with the same stationary distribution $\pi$ is not necessarily an HKY85 matrix).
\end{remark}

\begin{exmp}\rm
The well-known Kimura models with 2 or 3 parameters \cite{Kimura1980, Kimura1981} and the Jukes-Cantor model \cite{JC69} are also instances of ATR models. All these models have uniform stationary distribution $\pi$ and are $B$-time-reversible models with 
\begin{equation}\label{eq:fourier}
B=\left\{u^1=\frac{1}{4}\begin{pmatrix}
    1\\1\\1\\1
\end{pmatrix}, u^2=\frac{1}{4}\begin{pmatrix}
    1\\1\\-1\\-1
\end{pmatrix}, u^3=\frac{1}{4}\begin{pmatrix}
    1\\-1\\1\\-1
\end{pmatrix},u^4=\frac{1}{4}\begin{pmatrix}
    1\\-1\\-1\\1
\end{pmatrix}\right\}.    
\end{equation}

Working with coordinates in this basis simplifies the parametrization map, as already noted by \cite{Evans1993}. For these models this technique is also known as a discrete Fourier transform or Hadamard transform, see \cite{Hendy1994} for instance.  
 \end{exmp}

\subsection{Phylogenetic algebraic varieties}\label{sec:alg_var} 

As above, let $\pi$ be a fixed positive stationary distribution and let $B=\{u^1=\pi,\dots,u^{\kappa}\}$ be a $\pi$-orthogonal eigenbasis. We call $A$ the matrix of change of basis from $B$ to the standard basis $e^1,\dots, e^{\kappa}$.

\begin{remark}\label{rmk:reroot}\rm (Re-rooting) If we have a $B$-time-reversible  model on a phylogenetic tree $T$, one can chose any node of $T$ to play the role of the root and use the same transition matrices to describe the Markov process on $T$.  Indeed, let us prove that we can change the root from node $r$ to an adjacent node $s$ without changing transition matrices. Let $e_0$ be the edge from $r$ to $s$. Expression \eqref{eq:full} can be written as
\[ p_{\bm{i}}=\pi_{i_r}{M}^{e_0}_{i_r,i_s}\prod_{e \in E(T),e\neq e_0}M^e_{i_{p(e)},i_{c(e)}}.
\]
As $M^{e_0}$ is $\pi$-time-reversible, $\pi_{i_r}{M}^{e_0}_{i_r,i_s}=\pi_{i_s}{M}^{e_0}_{i_s,i_r}$, so expression \eqref{eq:full} is still valid if we root the tree at $s$.

From now on we do not specify the placement of a root (we conveniently place a root at any node if necessary). However, 
the distribution $\pi$ at the root node $r$ in expression \eqref{eq:full} is necessary: we cannot incorporate this distribution into one of the transition matrices on the edges adjacent to $r$ (as is usually done in the general Markov model or for $G$-equivariant models) because by doing so we would get a new matrix not belonging to the ATR model.
 

\end{remark}

As we are considering a $B$-time-reversible model, we have $\pi^r=\pi$ 
and the entries of $\pi^r$ are not free parameters anymore in the expression \eqref{eq_mapphi}. 
Thus, by extending to the complex numbers field, the map $\phi_T$ is defined on the parameter set 
\[\mathcal{P}^{\CC}= 
\{(M^e)_{e\in E(T)} \, \mid \, M^e  \in \mathcal{M}^1_{\CC}, (M^e)^t= A D^e A^{-1}   \},\]
where each $D^e$ is a diagonal matrix whose first entry is $1$, 
and the map $\phi_T$ is
\[\begin{array}{rcl}
    \mathcal{P}^{\CC}& \stackrel{\phi_T}{\longrightarrow} &\bigotimes^n W  \\
     \left( M^{e}\right)_{e \in E(T)} & \mapsto  & \sum\limits_{i_1,\dots,i_n} p_{i_1\dots i_n}^T  e^{i_1}\otimes \dots \otimes e^{i_n} \, ,
\end{array}\]
where $W=\langle e^1,\dots,e^{\kappa} \rangle_{\CC}$ is a $\CC$-vector space.

\begin{definition} The \emph{phylogenetic variety} of a tree $T$ evolving under a $B$-time-reversible model is the Zariski closure $\mathcal{V}_T$ of $\im  \phi_T$ in the tensor space $\otimes^n W$. We denote by $\mathcal{I}_T \subset \CC[p_{1\dots 1},\dots,p_{\kappa \dots \kappa}]$ the ideal of this algebraic variety. Its elements are called \emph{phylogenetic invariants}.
\end{definition}

The main goal of this work is to give phylogenetic invariants for ATR-models. As ATR-models are submodels of the general Markov model, phylogenetic invariants for the general Markov model are also in $\mathcal{I}_T$. Thus, Theorem \ref{thm:ARflat} holds and the $(\kappa+1)\times(\kappa+1)$ minors of those flattening matrices are phylogenetic invariants. 


\begin{remark}\label{rm:deg2}\rm (Degree two nodes)
Assume that $T$ has a degree two node $s$ and let $M^{e_1}$ and $M^{e_2}$ be transition matrices at the edges incident to it. Then the image by $\phi_T$ of these parameters coincides with the one obtained by deleting node $s$, joining $e_1$ and $e_2$ in a new edge $e_0$, and  considering the matrix $M^{e_0}=M^{e_1}M^{e_2}$ at $e_0$.
Therefore, adding or removing degree two nodes in a tree will not affect the map $\phi_T$ (when we add a degree two node on an edge and split it into two edges, we can trivially put the identity matrix at one of these edges).
\end{remark}

The map $\phi_T$ parametrizes a dense subset of $\mathcal{V}_T$. According to the result of \cite{chang1996} and its generalization in \cite{allman2004b}, if $T$ has no nodes of degree two, the fibers of $\phi_T$ are finite. Therefore the dimension of $\mathcal{V}_T$ coincides with the dimension of the space of parameters, which is $(\kappa-1)|E(T)|$ in this case. 

A special point in  $\im \phi_T$ is the image of identity transition matrices. This is called the \emph{no-evolution point} in \cite{CFM} and  is denoted as $p^0=\phi_T(\{Id\}_{e\in E(T)}).$ This point has a special relevance: in biological applications, transition matrices should not be far from identity (because it is difficult to obtain reliable data evolving on a tree with transition matrices far from the identity), so the points in $\mathcal{V}_T$ of most interest (biologically speaking) are those close to $p^0$.

In terms of probabilities it is easy to see that the image of $\phi_T$ lies in the hyperplane 
\begin{equation}\label{eq:H}
    H=\left\{p \in \otimes^n W \mid \, \sum_{\bm{i}}p_{\bm{i}}=1\right\},
\end{equation}
so we also have $\mathcal{V}_T\subset H$. Thus, $\sum_{\bm{i}}p_{\bm{i}}-1$ is a (trivial) phylogenetic invariant.


We make the previous map homogeneous by extending it to square matrices that diagonalize through $A$, without imposing $\lambda_1=1$. Let $C\mathcal{V}_T$ be the cone over $\mathcal{V}_T$, then $C\mathcal{V}_T$ is the Zariski closure of the following homogeneous map: 
\[\begin{array}{rcl}
    \tilde{\mathcal{P}}^{\CC} & \stackrel{\psi_T}{\longrightarrow} &\bigotimes^n W  \\
     \left( M^{e}\right)_{e \in E(T)} & \mapsto  & \sum\limits_{i_1,\dots,i_n} p_{i_1\dots i_n}^T  e^{i_1}\otimes \dots \otimes e^{i_n} \, ,
\end{array}\]
where $\tilde{\mathcal{P}}^{\CC}= \{(M^e)_{e\in E(T)} \, \mid \,  (M^e)^t= A D^e A^{-1}\}$ 
and $p_{i_1,\dots,i_n}^T$ is defined by the same expression as \eqref{eq:prob}.
Actually, $\mathcal{V}_T=C\mathcal{V}_T\cap H$ (see also \cite{Allman2008,CFK}). Indeed, if the row sum of a matrix $M^e$ is $\lambda_1^e$ (see Lemma \ref{lem:rowSum}), then $M^e$ is the product of $\lambda_1^e$  by a Markov matrix $\tilde{M}^e$ whose rows sum to one. Hence in \eqref{eq:full} we have $p_{\textbf{i}}^T=\left( \prod_{e\in E(T)} \lambda_1^e\right)\pi_{i_r}\prod_{e\in E(T)} \tilde{M}^e_{i_{p(e),c(e)}},$ and if $s:=\prod_{e\in E(T)} \lambda_1^e$, we have $\psi_T(\{M^e\})=s\phi_T(\{\tilde{M}^e\}).$
We extend the definition of a $B$-time-reversible model on a phylogenetic tree (\Cref{def:evol}) in order to allow all matrices $M^e$ to be  in $\mathcal{\tilde{P}}^{\CC}$.


We could do a change of coordinates in the parameter space $ \tilde{\mathcal{P}}^{\CC} $: instead of dealing with the entries of  $M^e$ we could deal directly with its eigenvalues $\lambda_1^e,\cdots,\lambda_{\kappa}^e.$ Thus, we could also express  $p_{i_1,\dots,i_n}^T $ in terms of the eigenvalues of $M^e$'s. This change in the parameter space and the analogous change of coordinates in the target  space will be studied in the next section.

\section{New coordinates for phylogenetic varieties of ATR models}\label{sec:coordinates}


From now on, the $\pi$-inner product will be simply denoted as $\langle \, , \, \rangle$. 
This inner product was introduced on $W=\RR^n$ but can be extended naturally to any tensor power $\otimes^n W= W\otimes \dots \otimes W$ as
\[\pe{w_1\otimes \dots \otimes w_n}{v_1\otimes \dots \otimes v_n}=\pe{w_1}{v_1}\, \pe{w_2}{v_2}\,\dots \,\pe{w_n}{v_n}. \]

Actually, it can also be extended to the complex number field by taking the conjugate of the second component in the inner product. However, we do not introduce this notation because in all inner products we will use, the second component will be a vector with real coordinates. Thus, we can think of $\pe{w}{v}$ with the definition we have already introduced and take $w$ in $\otimes^n \CC^{\kappa}$ and $v \in \otimes^n \RR^{\kappa}.$


Let $B=\{u^1=\pi,\dots, u^{\kappa}\}$ be a $\pi$-orthogonal eigenbasis. 
Then the basis 
$$B_n=\{u^{i_1}\otimes u^{i_2}\dots \otimes u^{i_n} \mid i_j \in \Sigma\}$$
is a $\pi$-orthogonal basis of $\otimes^n W.$ 
To simplify notation we call $$u^{\bm{i}}=u^{i_1}\otimes\dots \otimes u^{i_n} \, \textrm{ and } \, e^{\bm{i}}=e^{i_1}\otimes\dots \otimes e^{i_n} $$ if $\bm{i}=(i_1,\dots,i_n)\in \Sigma^{n}.$
Let $A$ be the $\kappa\times\kappa$ matrix of change of basis from $B$ to the standard basis as in the previous section. 

If $p \in \otimes^n W$ and $p_{i_1\dots i_n}$ are its coordinates in the standard basis, then its coordinates in the basis $B_n$ shall be denoted as $\bar{p}$ and are obtained as
$\bar{p}=(A^{-1}\otimes\dots\otimes A^{-1})\, p,$
where $\otimes$ denotes the Kronecker product of matrices.
That is, for ${\bm{i}}=(i_1,\dots,i_n)$, the  $\bm{i}$-coordinate of the tensor $p$ in the basis $B_n$, $\bar{p}_{i_1\dots i_n}$, is the  $i_1\dots i_n$ entry of the vector $\bar{p}$. 
Since $B_n$ is a $\pi$-orthogonal basis, this coordinate  $\bar{p}_{i_1\dots i_n}$ can also be computed as
\begin{equation}\label{eq:coord}
    \bar{p}_{i_1\dots i_n}=\frac{\pe{p}{u^{\bm{i}}}}
    {\pe{u^{\bm{i}}}{u^{\bm{i}}}} \, . 
\end{equation}



\noindent\textbf{Reparametrization.} Let $T$ be a tree evolving under a $B$-time-reversible model. If we change coordinates on the parameter space $\tilde{\mathcal{P}}^{\CC}$ so that (transposes of) transition matrices are diagonalized (and hence written in the basis $B$) and use coordinates $\bar{p}$ in the target space of $\psi_T$, we have a much simpler parametrization of $C\mathcal{V}_T$:

\[\begin{array}{rcl}
    \prod_{e\in E(T)}{\CC}^{\kappa} 
    & \stackrel{\varphi_T}{\longrightarrow} &\bigotimes^n W  \\
     \left( \Lambda^{e}\right)_{e \in E(T)} & \mapsto  & \sum\limits_{i_1,\dots,i_n} \bar{p}_{i_1\dots i_n}^T  u^{i_1}\otimes \dots \otimes u^{i_n} \, ,
\end{array}\]
where $\Lambda^{e}=\diag(\lambda^e_1,\dots, \lambda^e_{\kappa})$ is the diagonal matrix formed by the eigenvalues of $M^e$, $e \in E(T)$. We denote the Zariski closure of the image of $\varphi_T$ by $CV_T$.

\begin{definition}
We denote by ${I}_T$ the ideal of $CV_T$ in $\CC[\bar{p}_{1\dots 1},\dots,\bar{p}_{\kappa\dots \kappa}]$. 
A polynomial that belongs to all ${I}_T$ for $T \in \mathcal{T}_n$ is called a \emph{model invariant} (as it holds for any tree evolving under the $B$-time-reversible model). A polynomial that belongs to some ${I}_T$ for $T\in \mathcal{T}_n$ but that does not belong to ${I}_{T'}$ for some $T' \in \mathcal{T}_n$ is called a \emph{topology invariant}, see \cite[chapter 8]{SteelPhylogeny}.
\end{definition}
 
From the computational point of view, if we want to work with any $\pi$, we can work with polynomials with coefficients in the field of fractions $\CC(\pi_1,\dots,\pi_{\kappa}).$ Our computations in \texttt{Macaulay2} \cite{M2} follow this approach.

Using these new coordinates $\Bar{p}$ in the basis $B_n$, the following result relating marginalization and new coordinates will be useful. 
\begin{lemma}[Marginalization]\label{lem:marginal}
    For any $p\in \otimes^nW$ define the marginalization $p^+\in\oplus^{n-1}W$ of $p$ over the last component as
    \begin{equation}
        p^+_{i_1\dots i_{n-1}}= \sum_{j\in \Sigma} p_{i_1\dots i_{n-1}j}.
    \end{equation}
    Then, for a $\pi$-orthogonal basis $B_n$ as above, we have
    \begin{equation}
        \bar{p}_{i_1\dots i_{n-1}1}=\overline{p^+}_{i_1\dots i_{n-1}}.  
    \end{equation}
    Furthermore, if $T_n$ is an $n$-leaf tree and $T_{n-1} $ is the tree obtained from $T_n$ by deleting the pendant edge leading to leaf $n$, then for any $p\in \im(\phi_{T_n})$ we have that $p^+\in\im(\phi_{T_{n-1}}).$
\end{lemma}

\begin{proof}
    Note that the $(i,j)$-entry of $A^{-1}$ is the $i$-th coordinate of $e^j$ in the basis $B$, which is $\frac{\pe{e^j}{u^i}}{\pe{u^i}{u^i}}$. As $\pe{u^1}{u^1}=1$ and $\langle{e^i},{u^1}\rangle=1$ for any $i=1,\dots,\kappa$, the first row of $A^{-1}$ is $\mathbf{1}^t$. Thus, if $p\in \otimes^nW$, the slice of $\bar{p}$ with last component indexed by 1 is 
    \[(A^{-1}\otimes \stackrel{n-1}{\dots} \otimes A^{-1} \otimes \mathbf{1}^t)\, p,\]
    which is equal to $(A^{-1}\otimes \stackrel{n-1}{\dots} \otimes A^{-1})\, p^+.$ Thus, 
      $\bar{p}_{i_1\dots i_{n-1}1}=\overline{p^+}_{i_1\dots i_{n-1}}$.  
The last claim is well known and follows directly from \cite[Prop. 5.52]{Piotrbook}.
\end{proof}

\noindent\textbf{Markov action.} The following action of 
$GL_{\kappa}(\CC)^n$
on tensors in $\otimes^n W$
\[(N_1,\dots, N_n)\cdot p = (N_1\otimes \ldots \otimes N_n) p\]
is called the \emph{Markov action}. We can restrict this action to diagonal matrices so that we have an action of an $n\kappa$-dimensional torus 
$\mathbb{T}=(\CC^{\ast})^{\kappa} \times \stackrel{n}{\dots}\times (\CC^{\ast})^{\kappa}$. 
The torus $\mathbb{T}$ acts on tensors with coordinates $\bar{p}$ as follows: if $(D^1,\dots,D^n)$ is in $\mathbb{T}$ and $D^i =\diag(d^i_1,\dots,d^i_{\kappa} )$, then $(D^1,\dots,D^n)\cdot \bar{p}$ has coordinates $d^1_{i_1}\dots d^n_{i_n}\bar{p}_{i_1\dots i_n}.$ 

If $\bar{p}=\varphi_T((\Lambda^e)_{e\in E(T)})$, then  
$(D^1,\dots,D^n)\cdot \bar{p}=\varphi_T((\tilde{\Lambda}^e)_{e\in E(T)}),$
where $\tilde{\Lambda}^{e_i}=D^i\Lambda^{e_i}$ if $e_i$ is the pendant edge to leaf $l_i$, and $\tilde{\Lambda}^{e}=\Lambda^e$ otherwise. 
Hence we have that the Zariski closure of $\mathbb{T}\cdot CV_T$ is again 
$CV_T$, that is, $CV_T$ is invariant by the action of $\mathbb{T}$.



\begin{remark}\label{rem_action}
\rm Let $\bar{p} = \varphi_T(\{\Lambda^e\}_e)$. If $\bar{q}\in \im \varphi_T$ has the same parameters as $\bar{p}$ except the matrices $\Lambda^{e_i}$ on the pendant edges are replaced by identity matrices, then 
\begin{equation}\label{eq:id} 
\bar{p}=(\Lambda^{e_1},\dots,\Lambda^{e_n})\cdot \bar{q}, \quad \textrm{i.e. } \bar{p}_{i_1\dots i_n}=\lambda^{e_1}_{i_1}\dots\lambda^{e_n}_{i_n}\bar{q}_{i_1\dots i_n}.
\end{equation} 
Throughout the paper $\bar{q}$ will denote the image by $\varphi_T$ of a set of parameters with identity matrices at the pendant edges. 
\end{remark}

\subsection{Star trees evolving under ATR models}

In the following lemma we prove that if $T$ is a star tree, then $CV_T$ is a toric variety.
\begin{lemma}\label{lem:monomial} Let $T$ be the star tree with $n$ leaves and let it evolve under a $B$-time-reversible model. Then $CV_T$ is a toric variety (not necessarily normal), $\varphi_T$ is a monomial parametrization and $I_T\subset \CC[\bar{p}_{i_1\dots i_n} \,\mid \, i_j\in \Sigma]$ is generated by binomials. Moreover the no-evolution point $p^0$ is a non-singular point of $C\mathcal{V}_T$ and $\mathcal{V}_T$.

\end{lemma}
\begin{proof} 
By Remark \ref{rem_action}, we know that $CV_T$ is the closure of the orbit of ${p^0}=\varphi_T(\{Id\})$ under the action of $\mathbb{T}.$ This implies that $CV_T$ (and hence $C\mathcal{V}_T$) is a toric variety and ${p^0}$ is a non-singular point of $C\mathcal{V}_T$ and $\mathcal{V}_T$.
Again from Remark \ref{rem_action} we have that the parametrization $\varphi_T$ is monomial 
on the eigenvalues of $\Lambda^i$, given that the coordinates of $p^0$ in the basis $B_n$ are expressions in terms of $\pi$  only. 
From this, we obtain that the ideal ${I}_T$ can be generated by binomials (see \cite{ES96}).

\end{proof}


{For $G$-equivariant models it was already known that no-evolution points are non-singular points of star trees \cite[Corollary 3.9]{CFM}. The proof of \cite[Theorem 5.4]{CFM} shows that $p^0$ is a non-singular point on \emph{any} tree evolving under a $G$-equivariant model.}

\subsection{Gluing trees}

We recall here a procedure to glue trees and substitution parameters that was introduced in \cite{Allman2008}. 

\medskip
\noindent\textbf{Gluing trees.} Let $T_1$ and $T_2$ be two phylogenetic trees with leaf set $\{l_1,\dots,l_m,s_1\}$ and $\{s_2, l_{m+1}
, \dots, l_n\}$ respectively. We call $T'$ the tree with leaf set $\{l_1,\dots,l_n\}$ obtained by identifying $s_1$ and $s_2$ in a node $s$. We then call $T=T_1{*}T_2$ the tree obtained by deleting this node $s$ and replacing the two edges $e_1,e_2$ incident to it by a single edge $e_0$, see~\Cref{fig:gluing}. We call ${\alpha}=\{l_1,\dots,l_m\}$ and $\beta=\{l_{m+1}
,\dots,l_n\}$, so that $T$ has leaf set $L=\alpha \cup \beta$.


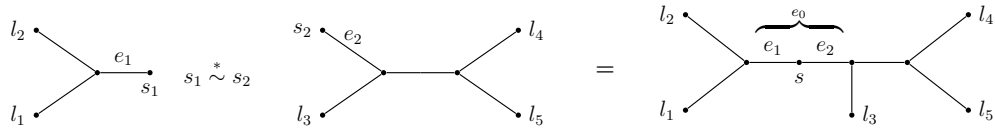
\begin{figure}[h]
\centering
\scalebox{0.7}{






\begin{forest}
for tree = {
    circle,
    fill=black,
    inner sep=1pt,
    edge = {semithick},
    l sep=7mm,
    s sep=15mm,
   }
[, coordinate, 
    [,  label=below:{$s_1$}, 
        no edge,
        edge label={node[midway]{$\overset{\ast}{s_1\sim s_2}$}},
        before computing xy={l=0, s=-25mm},
        [ , 
            edge label={node[midway, above]{$e_1$}},
            for tree={grow=west},
            before computing xy={l=0, s=-10mm},
            [,label=left:$l_2$]
            [,label=left:$l_1$]
        ] 
    ]
]
\end{forest}
\hspace{-0.2cm}
\begin{forest}
    for tree = {
        circle,
        fill=black,
        inner sep=1pt,
        edge = {semithick},
        l sep=7mm,
        s sep=15mm,
        }
        [,  coordinate,
            [ , 
                for tree={grow=west},
                before computing xy={l=0, s=-7mm},
                [, label=left:{$s_2$}, edge label={node[midway, above]{$e_2$}}]
                [,label=left:$l_3$]
            ]
            [ , 
                for tree={grow'=east}, 
                before computing xy={l=0, s=7mm},
                [,label=right:$l_4$]
                [,label=right:$l_5$]
            ]
        ]
\node at (current bounding box.east)[right=7ex] {\textbf{=}};
\end{forest}

\hspace{0.3cm}


\begin{forest}
for tree = {
    circle,
    fill=black,
    inner sep=1pt,
    edge = {semithick},
    l sep=7mm,
    s sep=17mm,
   }
[, label=below:{$s$}, 
    [ , 
        edge label={node[midway, above]{$e_1$}},
        for tree={grow=west},
        before computing xy={l=0, s=-10mm},
        [,label=left:$l_2$]
        [,label=left:$l_1$]{\node at (0,0.8) {$\overbrace{\quad\quad\quad \quad}^{e_0}$}; }
    ]   
    [ , 
        edge label={node[midway, above]{$e_2$}},
        for tree={grow'=east}, 
        before computing xy={l=0, s=10mm},
        [,  before computing xy={l=30, s=0mm},
            [,label=right:$l_4$]
            [,label=right:$l_5$]
        ]
        [,
            before computing xy={l=0, s=-10mm},
            label=right:$l_3$]
    ]
]
\end{forest}




 }
\caption{Gluing of a tripod and a quartet. The result is a tree with 5 leaves. In this case $m=2,n=5,\alpha=\{l_1,l_2\}$ and $\beta=\{l_3,l_4,l_5\}$.}
\label{fig:gluing}
\end{figure}


\medskip
\noindent\textbf{Gluing parameters}. 
If $T_1$ and $T_2$ evolve under the $B$-time-reversible model with matrices $(M^e)_{e\in E(T_1)\cup E(T_2)}$, then we define transition matrices at the edges of $T=T_1*T_2$ as follows: if $e\neq e_0$, then $e\in E(T_i) $ for some $i=1,2$ and  we assign to $e$ the transition matrix as in $T_i$; if $e=e_0$, we let $M^e=M^{e_1}M^{e_2}$.

According to Remark \ref{rm:deg2}, we can indistinguishably use the tree $T=T_1{*} T_2$ with the set of parameters  just described, or the tree $T'$ with the degree two node $s$ and parameters $(M^e)_{e\in E(T_1)\cup E(T_2)}$.

\medskip

The following lemma is obvious if we think about distributions, but 
for the sake of completeness 
we prove it 
in \Cref{app:gluing}
for any set of parameters.
\begin{lemma}\label{lem:Ys} 
Let $T$ be obtained by gluing two trees $T_1$ and $T_2$ as above, let $p^{T_i}\in \im \psi_{T_i}$, $i=1,2$, and let $p^T$ be the  tensor obtained by gluing parameters on $T_1$ and $T_2$. Let $\bm{i}=(\bm{i}_{\alpha},\bm{i}_{\beta})$ be a collection of states at the leaves of $T$. Then, for any state $k \in \Sigma$ at node $s=s_1\sim s_2$ we have
\[p^T_{\bm{i}_{\alpha},k,\bm{i}_{\beta}}=\frac{1}{\pi_{k}}p_{\bm{i}_{\alpha},k}^{T_1}p_{k,\bm{i}_{\beta}}^{T_2}.\]   

\end{lemma}

The following theorem expresses a tensor evolving on $T=T_1*T_2$ under a $B$-time-reversible model in terms of tensors evolving on $T_i$. Here $\flat(\bar{p})$ is as in Definition \ref{def:flat} exchanging coordinates in the standard basis by coordinates in $B_n$.  
\begin{thm}\label{thm:gluing}
    Let $T_1$ and $T_2$ be two trees evolving under a $B$-time reversible model and let $T=T_1* T_2$  be the tree obtained by gluing $T_1$ and $T_2$  as above. Let $p^{T_i}\in \im \psi_{T_i}$, $i=1,2$ and let $p^T$ be the  tensor obtained by gluing parameters on $T_1$ and $T_2$. 
    Then, in coordinates in the basis $B_n$, we have
\begin{equation}\label{eq:gluing}
    \bar{p}_{i_1\dots i_n}=\sum_{j\in \Sigma} \langle u^j,u^j\rangle
   \, \bar{p}^{T_1}_{i_1...i_m j}\bar{p}^{T_2}_{j i_{m+1}\dots i_n},
    \end{equation}
for any $i_1,\dots,i_n\in \Sigma.$
If $B$ is a $\pi$-orthonormal eigenbasis, then the expression becomes
\begin{equation}
    \bar{p}_{i_1\dots i_n}=\sum_{j\in \Sigma} 
   \, \bar{p}^{T_1}_{i_1...i_m j}\bar{p}^{T_2}_{j i_{m+1}\dots i_n}
    \end{equation}
 and we have 
 $\flat_{1\dots m|m+1\dots n}(\bar{p})=  \flat_{1\dots m|s_1}(\bar{p}^{T_1})\flat_{s_2|m+1\dots n}(\bar{p}^{T_2})$.
\end{thm}

As an immediate  consequence of the last statement we recover Theorem \ref{thm:ARflat} for ATR models. 
We proceed to prove the theorem.
\begin{proof}
 Let $\bm{i}=(i_1,\dots,i_n)=(\bm{i}_{\alpha},\bm{i}_{\beta})$ be a collection of states at the leaves of $T$. 
 We start by expressing $\langle p, u^{\bm {i}} \rangle$ in terms of scalar products of the corresponding tensors on the subtrees $T_1$ and $T_2$.
    We have
    \[
     \pe{p}{u^{\bm{i}}}=\left\langle\sum_{\bm{j}=(\bm{j}_{\alpha},\bm{j}_{\beta})} p_{\bm{j}} e^{\bm{j}},u^{\bm{i}}\right\rangle=\sum_{\bm{j}_{\alpha},\bm{j}_{\beta}}p_{\bm{j}_{\alpha},\bm{j}_{\beta}} \pe{ e^{\bm{j}_{\alpha}}}{u^{\bm{i}_{\alpha}}} \pe{ e^{\bm{j}_{\beta}}}{u^{\bm{i}_{\beta}}}.
     \]
By \eqref{eq:prob},  
$p_{\bm{j}_{\alpha},\bm{j}_{\beta}}=\sum_{j_s \in \Sigma}p_{\bm{j}_{\alpha},j_s,\bm{j}_{\beta}}$
where $s=s_1\sim s_2$, and by Lemma \ref{lem:Ys} 
we get

\[p_{\bm{j}_{\alpha},\bm{j}_{\beta}}
=\sum_{j_s\in \Sigma} \frac{p^{T_1}_{\bm{j}_{\alpha},j_s}}{\pi_{j_s}}p^{T_2}_{j_s,\bm{j}_{\beta}}= \sum_{j_r,j_s\in \Sigma} \frac{p^{T_1}_{\bm{j}_{\alpha},j_s}}{\pi_{j_s}} \delta_{j_r,j_s} p^{T_2}_{j_r,\bm{j}_{\beta}},\]
where $\delta_{i,j}$ is the Kronecker delta.
Hence,
\begin{equation*}
 \langle p, u^{\bm{i}} \rangle = \sum_{\bm{j}_{\alpha},j_s} \frac{p^{T_1}_{\bm{j}_{\alpha},j_s}}{\pi_{j_s}} \langle e^{\bm{j}_{\alpha}},u^{\bm{i}_{\alpha}} \rangle \left( \sum_{j_r} \delta_{j_s,j_r} \sum_{\bm{j}_{\beta}}p^{T_2}_{j_r,\bm{j}_{\beta}}\langle e^{\bm{j}_{\beta}},u^{\bm{i}_{\beta}} \rangle\right).
\end{equation*} 
Now we observe that 
\begin{equation*}
\delta_{i,j}=\frac{1}{\pi_i}\pi_i(e^i)^t  e^j=\pe{\pi_i e^i}{e^j}= 
\left\langle{\pi_ie^i,\sum_{k\in \Sigma} \frac{\pe{e^j}{u^k}}{\pe{u^k}{u^k}}u^k}\right\rangle=\sum_{k\in \Sigma} \frac{\pi_{i} \, 
\pe{e^i}{u^k} \, \pe{u^k}{e^j}}{\pe{u^k}{u^k}}.  
\end{equation*}
Therefore,  using this expression we obtain
\begin{equation*}
\begin{split}
\langle p, u^{\bm{i}} \rangle &= 
 \sum_{\bm{j}_{\alpha},j_s} \frac{p^{T_1}_{\bm{j}_{\alpha},j_s}}{\pi_{j_s}} \langle e^{\bm{j}_{\alpha}},u^{\bm{i}_{\alpha}} \rangle \left(\sum_{j_r}\sum_{k\in \Sigma}  \frac{\pi_{j_s} \, 
\pe{e^{j_s}}{u^k} \, \pe{u^k}{e^{j_r}}}{\pe{u^k}{u^k}} \sum_{\bm{j}_{\beta}}p^{T_2}_{j_r,\bm{j}_{\beta}}\langle e^{\bm{j}_{\beta}},u^{\bm{i}_{\beta}} \rangle \right)\\
&= \sum_{\bm{j}_{\alpha},j_s} \frac{p^{T_1}_{\bm{j}_{\alpha},j_s}}{\pi_{j_s}} \langle e^{\bm{j}_{\alpha}},u^{\bm{i}_{\alpha}} \rangle \left(\sum_{k\in \Sigma}  \frac{\pi_{j_s} \, 
\pe{e^{j_s}}{u^k}}{\pe{u^k}{u^k}} \sum_{j_r,\bm{j}_{\beta}}p^{T_2}_{j_r,\bm{j}_{\beta}}\pe{u^k}{e^{j_r}}
\langle e^{\bm{j}_{\beta}},u^{\bm{i}_{\beta}} \rangle\right).
\end{split} 
\end{equation*}
The last sum in the previous expression is $\langle p^{T_2},u^{k}\otimes u^{\bm{i}_{\beta}}\rangle $  so we get
\begin{equation*}
\langle p, u^{\bm{i}} \rangle
=\sum_{k \in \Sigma} \frac{1}{\pe{u^k}{u^k}}
\sum_{\bm{j}_{\alpha},j_s} p^{T_1}_{\bm{j}_{\alpha},j_s}\langle e^{\bm{j}_{\alpha}},u^{\bm{i}_{\alpha}} \rangle  
\pe{e^{j_s}}{u^k}
\langle p^{T_2},u^{k}\otimes u^{\bm{i}_{\beta}}\rangle \, .
\end{equation*}
The proof finishes by observing that $\langle p^{T_1}, u^{\bm{i}_{\alpha}} \otimes u^k\rangle =
\sum_{\bm{j}_{\alpha},j_s} p^{T_1}_{\bm{j}_{\alpha},j_s}\langle e^{\bm{j}_{\alpha}},u^{\bm{i}_{\alpha}} \rangle  
\pe{e^{j_s}}{u^k}$
and dividing $\langle p,  u^{\bm{i}} \rangle$ by $\langle  u^{\bm{i}},  u^{\bm{i}} \rangle$:

{\small 
\begin{equation*}
\bar{p}_{\bm{i}} =\frac{\langle p,  u^{\bm{i}} \rangle}{\langle  u^{\bm{i}},  u^{\bm{i}} \rangle}
=
\sum_{k} \langle u^k, u^k \rangle \frac{\langle p^{T_1}, u^{\bm{i}_{\alpha}} \otimes u^k\rangle}{ \langle u^{\bm{i}_{\alpha}},u^{\bm{i}_{\alpha}} \rangle \langle u^k, u^k\rangle} \frac{\langle p^{T_2},  u^k \otimes u^{\bm{i}_{\beta}} \rangle}{\langle u^k, u^k\rangle \langle u^{\bm{i}_{\beta}}, u^{\bm{i}_{\beta}} \rangle} =\sum_{k \in \Sigma} \langle u^k, u^k \rangle \, \bar{p}^{T_1}_{\bm{i}_{\alpha}k}\, \bar{p}^{T_2}_{k \bm{i}_{\beta}}. \\
\end{equation*} }
The last claim in the statement of the theorem is straightforward.
\end{proof}

The above theorem is also valid when one of the two subtrees, say $T_2$, is a tree with two leaves and a single edge. In this case this operation is equivalent to multiplying $\bar{p}^{T_1}$ by a diagonal matrix and so it is equivalent to the Markov action on one leaf.

\begin{remark}
In general, the gluing procedure is not equivalent to a toric fiber product  as described in \cite{Sullivant2007}. Indeed, as we will see in section 5.3 for the TN93 model, the parameterization for quartets obtained by gluing two tripods (which have a monomial parameterization) is not monomial and hence the corresponding ideal is not the toric fiber product of both toric ideals. However, for the toric models Kimura 3-parameter and its submodels studied in \cite{Sturmfels2005}, the gluing procedure is equivalent to a toric fiber product if we use Fourier coordinates.    
\end{remark}

Thus, with Theorem \ref{thm:gluing} we recover a well known result of Evans and Speed \cite{Evans1993}.  

\begin{cor}(\cite{Evans1993,Sturmfels2005})
On a tree evolving under the Kimura 3-parameter model or one of its submodels (Kimura 2-parameter and Jukes-Cantor), the Fourier coordinates of a  tensor $p\in \im \psi_T$ have a monomial expression in terms of the eigenvalues of the transition matrices. Moreover, the ideal of the corresponding phylogenetic variety $V_T$ is generated by binomials.
\end{cor}

\begin{proof}
    For the Kimura 3-parameter model and its submodels, $\pi$ is the uniform distribution. The Fourier basis in \eqref{eq:fourier} 
    is a $\pi$-orthonormal basis that diagonalizes all transition matrices in these models. If we biject the set $\Sigma=\{1,2,3,4\}$ with the additive group $G=\mathbb{Z}_2\times \mathbb{Z}_2$ by identifying $1 = (0,0), 2=(0,1), 3=(1,0),4=(1,1)$, and denote by $\boxplus$ 
    the sum  in $G$, then it is easy to see that for a tripod tree we have (see also \cite{Sturmfels2005})
    \[\bar{p}_{i_1i_2i_3}=\left\{ \begin{array}{cc}
       \lambda^1_{i_1}\lambda^2_{i_2}\lambda^3_{i_3},  &  \textrm{ if } i_1\boxplus i_2 \boxplus i_3=0; \\
       0,  & \textrm{ otherwise}.
    \end{array}\right. \]
    Then the corollary follows easily by Theorem \ref{thm:gluing} and induction. Indeed, if $T$ is an $n$-leaf tree, we consider a cherry on it and view $T$ as $T=T_1*T_2$, where $T_2$ is a tripod tree. The induction hypothesis is that for any $m<n$, $\bar{p}_{i_1\dots i_m}=0$ if $i_1 \boxplus\dots \boxplus i_m\neq 0$ and $\bar{p}_{i_1\dots i_m}$ has a monomial expression in the eigenvalues if $i_1\boxplus\dots\boxplus i_m= 0$. By Theorem \ref{thm:gluing} we have 
     \begin{equation*}
    \bar{p}_{i_1\dots i_n}=\sum_{j\in \Sigma} 
   \, \bar{p}^{T_1}_{i_1...i_{n-2} j}\bar{p}^{T_2}_{j i_{n-1} i_n}
    \end{equation*}
so that, by the induction hypothesis, the only (possibly) non-zero summand is for $j=i_{n-1} \boxplus i_n$, which implies $i_1 \boxplus\dots \boxplus i_n=0$ and gives a monomial expression.
\end{proof}

Theorem \ref{thm:gluing} gives an inductive procedure to build the parametrization map of a phylogenetic tree evolving under a $B$-time-reversible model. Although this gluing procedure mimics the one described in \cite{Allman2008} and \cite{Draisma}, it is not obvious that this leads to the expression of the ideal of these phylogenetic varieties in terms of subtrees as was done in these two papers. We pose the following question:

\begin{question}
Can the ideal of the phylogenetic variety evolving on a $B$-time-reversible model be described in terms of the ideals of flattenings at its interior nodes and edges as in \cite{Allman2008, Draisma}?
\end{question}

We do not deal with this general question in this paper, but for the TN93 model on quartets we construct from tripods and edge flattenings a local complete intersection that describes the variety on an open set (see \Cref{sec:TN93}). 







\section{Invariants for trees evolving under the TN93 model}\label{sec:TN93}

In this section we showcase how the framework developed previously can be used to find phylogenetic invariants for the Tamura-Nei model TN93 introduced in Example \ref{ex:TN93}. The stationary distribution $\pi$ is fixed (and we will assume it to be generic when convenient) and the $\pi$-orthogonal basis $B$ is specified in Example \ref{ex:TN93}.
 We start by computing phylogenetic invariants for trees with three leaves, then we use the marginalization and tree gluing to find phylogenetic invariants for quartets and $n$-leaf trees.
\subsection{Invariants for tripods}



Let $T$ be a star tree with three leaves $l_1,l_2,l_3$ and three edges $e_1,e_2,e_3$ (we call it the \emph{tripod}). 
The joint probability tensor for the point of no-evolution $p^0=\psi_T(Id,Id,Id)$ has the following coordinates in the standard basis:
\begin{equation}\label{eq:q}  p^0_{i_1i_2i_3}=\sum_{i_r\in\Sigma}\pi_{i_r}{Id}_{i_r,i_1}{Id}_{i_r,i_2}{Id}_{i_r,i_3}=\left\lbrace\begin{array}{ll}
   \pi_{i_r},  & \mbox{if }i_1=i_2=i_3=i_r;\\
    0, & \mbox{otherwise.}
\end{array}\right.
\end{equation}


\begin{lemma}\label{lem:tripodq} Consider the evolutionary model TN93 on the tripod $T$. The coordinates $\bar{p}^0$ in the basis $B_n$ for the no-evolution point $p^0$ are:
\begin{gather*}
 \bar{p}^0_{111}=1, \qquad \bar{p}^0_{222}=\frac{\pi_{34}-\pi_{12}}{\pi_{12}^2\pi_{34}^2}, \qquad \bar{p}^0_{333}=\frac{\pi_{4}^2-\pi_{3}^2}{\pi_{3}^2\pi_{4}^2}, \qquad \bar{p}^0_{444}=\frac{\pi_{2}^2-\pi_{1}^2}{\pi_{1}^2\pi_{2}^2},\\
\bar{p}^0_{122}=\bar{p}^0_{212}=\bar{p}^0_{221}=\frac{1}{\pi_{12}\pi_{34}}, \qquad \bar{p}^0_{133}=\bar{p}^0_{313}=\bar{p}^0_{331}=\frac{\pi_{34}}{\pi_{3}\pi_{4}}, \\
\bar{p}^0_{144}=\bar{p}^0_{414}=\bar{p}^0_{441}=\frac{\pi_{12}}{\pi_{1}\pi_{2}}, \qquad     \bar{p}^0_{233}=\bar{p}^0_{323}=\bar{p}^0_{332}=\frac{-1}{\pi_{3}\pi_{4}},\\
\bar{p}^0_{244}=\bar{p}^0_{424}=\bar{p}^0_{442}=\frac{1}{\pi_{1}\pi_{2}}, \quad \mbox{ and } \quad \bar{p}^0_{i_1i_2i_3}=0 \mbox{ otherwise.}
\end{gather*}

\end{lemma}

\begin{remark}\label{rem:tripod_lin_inv}\rm 
    For a distribution $\pi$ such that  $\pi_{12}\neq \pi_{34}$, $\pi_1\neq \pi_2$, and $\pi_3\neq\pi_4$, there are exactly 45 entries of $\bar{p}^0_{i_1i_2i_3}$ that vanish: when $\{i_1,i_2,i_3\}$ contains either a unique $3$ or a unique $4$ or when $i_j=2$ for some $j$ and $i_k=1$ for $k\neq j$.
    These are the genericity conditions we will consider for the distribution $\pi$ from now on. 
\end{remark}
\begin{proof}[Proof of~\Cref{lem:tripodq}]
From \eqref{eq:coord} and \eqref{eq:q} we have the expression


\begin{equation*}
\bar{p}^0_{i_1i_2i_3}=\frac{1}{\pe{u^{\mathbf{i}}}{u^{\mathbf{i}}}}\sum_{j\in\Sigma}\pi_j\pe{e^j}{u^{i_1}}\pe{e^j}{u^{i_2}}\pe{e^j}{u^{i_3}},   
\end{equation*}
which is invariant after reordering $i_1,i_2,i_3$. First we prove that $\bar{p}^0_{i_1i_2i_3}=0$ for the cases mentioned in the previous remark. 



\underline{Case I}: $\{i_1,i_2,i_3\}$ contains only one $3$.
Without loss of generality assume that $i_1=3$ and $i_2,i_3\neq 3$. From~\eqref{eq:table_scalar} we have
\begin{equation*}
\begin{split}
    \bar{p}^0_{3i_2i_3}=&\frac{1}{\pe{u^{\mathbf{i}}}{u^{\mathbf{i}}}}\sum_{j\in\Sigma}\pi_j\pe{e^j}{u^{3}}\pe{e^j}{u^{i_2}}\pe{e^j}{u^{i_3}}\\
    =&\frac{1}{\pe{u^{\mathbf{i}}}{u^{\mathbf{i}}}}\left(\pi_3\pe{e^3}{u^{3}}\pe{e^3}{u^{i_2}}\pe{e^3}{u^{i_3}} + \pi_4\pe{e^4}{u^{3}}\pe{e^4}{u^{i_2}}\pe{e^4}{u^{i_3}}\right)\\
    =&\frac{1}{\pe{u^{\mathbf{i}}}{u^{\mathbf{i}}}}\left(\frac{\pi_3\pi_4}{\pi_{34}}\pe{e^3}{u^{i_2}}\pe{e^3}{u^{i_3}} - \frac{\pi_3\pi_4}{\pi_{34}}\pe{e^4}{u^{i_2}}\pe{e^4}{u^{i_3}}\right)=0,
\end{split}
\end{equation*}
where the last equality holds because the last two rows of~\eqref{eq:table_scalar} for $u^i\neq u^3$ are equal. 

\underline{Case II}: $\{i_1,i_2,i_3\}$ contains only one $4$. The proof is analogous to the previous case by noting that the first two rows of~\eqref{eq:table_scalar} are equal for $u^i\neq u^4$.



\underline{Case III}: $i_j=2$ for some $j$ and $i_k=1$ for $k\neq j$: Assume, without loss of generality, that $i_1=i_2=1$ and $i_3=2$. We have 
\begin{equation*}
        \bar{p}^0_{112}=\frac{1}{\pe{u^{\mathbf{i}}}{u^{\mathbf{i}}}}\sum_{j\in\Sigma}\pi_j\pe{e^j}{u^{1}}\pe{e^j}{u^{1}}\pe{e^j}{u^{2}}
        = \frac{1}{\pe{u^{\mathbf{i}}}{u^{\mathbf{i}}}} \left( \pi_{12}\pi_{34}-\pi_{12}\pi_{34}\right)=0.
\end{equation*}



The remaining $64-45=19$ coordinates $\bar{p}^0_{i_1i_2i_3}$ are non-zero for generic $\pi$ and can be easily computed in a similar fashion.
\end{proof}


By \eqref{eq:id}, for any point $p\in \im \varphi_T$ 
    the coordinates $\Bar{p}_{i_1i_2i_3}$ with $i_1,i_2$ and $i_3$ satisfying any of the cases of~\Cref{rem:tripod_lin_inv} vanish. Thus, $CV_T$ is contained in a linear space $\mathcal{L}_3\subset \otimes^3W$ of dimension 19 defined by the 45 linear equations, and $V_T$ is contained in the linear space $\mathcal{L}_3\cap H$ of dimension 18, where $H$ is the space defined in~\eqref{eq:H}.


We now provide a set of generators for $I_T$ and a set of polynomials defining a complete intersection that cuts out $CV_T\subset \mathcal{L}_3$ in an open set.

\begin{prop}\label{lem:tripodp}
     Let $T$ be a tripod that evolves under the TN93 model.  For generic $\pi$, $I_T$ is a binomial ideal minimally generated by 45 linear monomials, 9 quadric binomials, 29 cubic binomials, and 3 quintic binomials. If we consider $CV_T\subset\mathcal{L}_3$, 
     then the variety $X_T$ defined by the 9 polynomials
    \begin{gather*}
\bar{p}_{222}\bar{p}_{441}-\frac{\p{34}-\p{12}}{\p{34}}\bar{p}_{221}\bar{p}_{442},\qquad
\bar{p}_{222}\bar{p}_{414}-\frac{\p{34}-\p{12}}{\p{34}}\bar{p}_{212}\bar{p}_{424},\\
\bar{p}_{222}\bar{p}_{144}-\frac{\p{34}-\p{12}}{\p{34}}\bar{p}_{122}\bar{p}_{244}, \qquad  \bar{p}_{332}\bar{p}_{441}+\frac{\p{12}}{\p{34}}\bar{p}_{331}\bar{p}_{442},\\
\bar{p}_{323}\bar{p}_{414}+\frac{\p{12}}{\p{34}}\bar{p}_{313}\bar{p}_{424}, \qquad
\bar{p}_{233}\bar{p}_{144}+\frac{\p{12}}{\p{34}}\bar{p}_{133}\bar{p}_{244},\\
\bar{p}_{144}\bar{p}_{414}\bar{p}_{441}-\frac{\p{1}\p{2}\p{12}}{(\p{1}-\p{2})^2}\bar{p}_{111}\bar{p}_{444}^2,  \qquad \bar{p}_{133}\bar{p}_{313}\bar{p}_{331}-\frac{\p{3}\p{4}\p{34}}{(\p{3}-\p{4})^2}\bar{p}_{111}\bar{p}_{333}^2,\\
\bar{p}_{332}\bar{p}_{323}\bar{p}_{233}-\frac{\p{3}\p{4}\p{12}^2}{(\p{3}-\p{4})^2(\p{12}-\p{34})}\bar{p}_{222}\bar{p}_{333}^2,
    \end{gather*}
    is a complete intersection that cuts out $CV_T$ in the open set $\bar{p}_{{i}{i}{i}}\neq 0$ for $i=1,2,3,4$. Furthermore, $CV_T$ is an irreducible component of $X_T$. 
\end{prop}

\begin{proof}
    From~\Cref{lem:monomial} it follows that $\varphi_T$ is a monomial map and $I_T$ is generated by binomials. \Cref{lem:tripodq} and \eqref{eq:id} guarantees that the 45 linear polynomials  $\bar{p}_{i_1i_2i_3}$ are in $I_T$ for the indices $\{i_1,i_2,i_3\}$ satisfying the conditions of~\Cref{rem:tripod_lin_inv}. \texttt{Macaulay2} computations provide the remaining generators and guarantee that they are a minimal set of generators. The comprehensive list is included in~\Cref{app:Computations}.

    To study $CV_T$, we use tools from toric geometry. We refer the reader to \cite[Chapter~4]{sturmfelsgröbner} for an introduction to toric ideals. Consider a rescaling $\tilde{p}$ of $\bar{p}$ such that 
    is a monic monomial parametrization. 
    The vanishing ideal of  ${V}_T$ is the toric ideal generated by the columns of the matrix $C$ in \eqref{eq:C_tripod}, 
    \begin{equation}\label{eq:C_tripod}
        C=\left(\begin{array}{cccc|ccc|ccc|ccc|ccc|ccc}
            1  &  0 & 0 & 0 & 1 & 0 & 0 & 1 & 0 & 0 & 1 & 0 & 0 & 0 & 0 & 0 & 0 & 0 & 0\\
            0  &  1 & 0 & 0 & 0 & 1 & 1 & 0 & 0 & 0 & 0 & 0 & 0 & 1 & 0 & 0 & 1 & 0 & 0\\
            0  &  0 & 1 & 0 & 0 & 0 & 0 & 0 & 1 & 1 & 0 & 0 & 0 & 0 & 1 & 1 & 0 & 0 & 0\\
            0  &  0 & 0 & 1 & 0 & 0 & 0 & 0 & 0 & 0 & 0 & 1 & 1 & 0 & 0 & 0 & 0 & 1 & 1\\ 
            \hline
            1  &  0 & 0 & 0 & 0 & 1 & 0 & 0 & 1 & 0 & 0 & 1 & 0 & 0 & 0 & 0 & 0 & 0 & 0\\
            0  &  1 & 0 & 0 & 1 & 0 & 1 & 0 & 0 & 0 & 0 & 0 & 0 & 0 & 1 & 0 & 0 & 1 & 0\\
            0  &  0 & 1 & 0 & 0 & 0 & 0 & 1 & 0 & 1 & 0 & 0 & 0 & 1 & 0 & 1 & 0 & 0 & 0\\
            0  &  0 & 0 & 1 & 0 & 0 & 0 & 0 & 0 & 0 & 1 & 0 & 1 & 0 & 0 & 0 & 1 & 0 & 1\\
            \hline
            1  &  0 & 0 & 0 & 0 & 0 & 1 & 0 & 0 & 1 & 0 & 0 & 1 & 0 & 0 & 0 & 0 & 0 & 0\\
            0  &  1 & 0 & 0 & 1 & 1 & 0 & 0 & 0 & 0 & 0 & 0 & 0 & 0 & 0 & 1 & 0 & 0 & 1\\
            0  &  0 & 1 & 0 & 0 & 0 & 0 & 1 & 1 & 0 & 0 & 0 & 0 & 1 & 1 & 0 & 0 & 0 & 0\\
            0  &  0 & 0 & 1 & 0 & 0 & 0 & 0 & 0 & 0 & 1 & 1 & 0 & 0 & 0 & 0 & 1 & 1 & 0\\
        \end{array}\right).
    \end{equation}

    The columns of the matrix $D$ in~\eqref{eq:D_tripod}  are generators of $\ker (C)$, hence the 9 polynomials defined in the statement 
    are the generators of the lattice basis ideal $J$ of $C$ when considering the aformentioned parametrization $\tilde{p}$.

    \begin{equation}\label{eq:D_tripod}
        D=\left(\begin{array}{ccc|ccc|ccc}
            0& 0& 0& 0& 0& 0& -1& -1& 0\\
            1& 1& 1& 0& 0& 0& 0& 0& -1\\
            0& 0& 0& 0& 0& 0& 0& -2& -2\\
            0& 0& 0& 0& 0& 0& -2& 0& 0\\
            \hline
            0& 0& -1& 0& 0& 0& 0& 0& 0\\
            0& -1& 0& 0& 0& 0& 0& 0& 0\\
            -1& 0& 0& 0& 0& 0& 0& 0& 0\\
            \hline
            0& 0& 0& 0& 0& -1& 0& 1& 0\\
            0& 0& 0& 0& -1& 0& 0& 1& 0\\
            0& 0& 0& -1& 0& 0& 0& 1& 0\\
            \hline
            0& 0& 1& 0& 0& 1& 1& 0& 0\\
            0& 1& 0& 0& 1& 0& 1& 0& 0\\
            1& 0& 0& 1& 0& 0& 1& 0& 0\\
            \hline
            0& 0& 0& 0& 0& 1& 0& 0& 1\\
            0& 0& 0& 0& 1& 0& 0& 0& 1\\
            0& 0& 0& 1& 0& 0& 0& 0& 1\\
            \hline
            0& 0& -1& 0& 0& -1& 0& 0& 0\\
            0& -1& 0& 0& -1& 0& 0& 0& 0 \\
            -1& 0& 0& -1& 0& 0& 0& 0& 0
        \end{array}\right)
    \end{equation}

    \noindent
    Theorem 2.1 in~\cite{CompInter_BinomialIdeals} (see also Theorem 2.7 in~\cite{RegSeq_BinomialIdeals}) guarantees that this lattice basis ideal is a complete intersection if every mixed submatrix of $D$ has at least as many columns as rows. Recall that a matrix is mixed if in every column there is at least one positive and one negative entry. From the structure of $D$ it is intuitive to see that every mixed submatrix is either a square matrix or has more rows than columns, however we provide a code that verifies that every submatrix of $D$ with fewer rows than columns is not mixed. The code is included in the  \href{https://dataverse.csuc.cat/dataset.xhtml?persistentId=doi:10.34810/data1128}{CORA repository}.
    By \cite[Corollary 2.5]{ES96} $I_T=J:(\prod_{ijk}\tilde{p}_{ijk})^\infty$, 
    where 
    $J:(\prod_{ijk}\tilde{p}_{ijk})^\infty$ denotes the saturation of $J$ by the ideal generated by the product of all variables, 
    and by~\cite[Corollary 2.1]{HOSTEN2000625}, $I_T$ is a minimal prime of $J$, hence $CV_T$ is an irreducible component of $X_T$. 
\end{proof}

\begin{remark}\rm 
    For a non-generic distribution $\pi$, there might be more invariants arising from the vanishing of $\Bar{p}^0_{iii}$, with $i\in\{2,3,4\}$. For instance, when $\pi_{12}=\pi_{34}$, $\pi_1=\pi_2$ and $\pi_3=\pi_4$, then $I_T$ is minimally generated by 9 quadrics, 24 cubics, 3 quintics and the additional linear invariants $\bpt{2}{2}{2}=\bpt{3}{3}{3}=\bpt{4}{4}{4}=0$. 

\end{remark}

\subsection{Invariants for \texorpdfstring{$n$}{n}-leaf trees arising from tripods}

In this section we focus on phylogenetic invariants of trees with $n$ leaves evolving under TN93 that can be obtained from invariants of the tripod either by gluing or by marginalization. 
Let $T$ be a tree with $n$ leaves, $n>3$, and consider a tensor ${p}=\varphi_T(\lbrace \Lambda^e\rbrace_{e\in E(T)})$. 
We can always obtain ${p}$ by gluing a tensor ${p}^{T_1}$ on a tripod $T_1$ to a tensor $p^{T_2}$ on an $(n-1)$-leaf tree $T_2$. Indeed, let $l_1$ and $l_2$ be two leaves in $T$ forming a cherry and consider the interior edge $e_0$ adjacent to the cherry; then we split this edge into two edges $e_1$ and $e_2$ as in \Cref{fig:gluing} and call $T_1$  the tripod tree formed by the cherry and $e_1$ and $T_2$ the tree $T\setminus T_1$. 
\Cref{thm:gluing} ensures that ${p}={p^{T_1}\ast p^{T_2}}$, where 
${p}^{T_1}=\varphi_{T_1}(\lbrace \Lambda^e\rbrace_{e\in E(T_1)})$ and
${p}^{T_2}=\varphi_{T_2}(\lbrace \Lambda^e\rbrace_{e\in E(T_2)})$,
with $\Lambda^{e_1}=\Lambda^{e_0}$, $\Lambda^{e_2}=Id$, where $e_0,e_1,e_2$ are the edges involved in the gluing as denoted in \Cref{fig:gluing}, and the remaining matrices $\Lambda^e$ for $e\in E(T_i)$ coincide with $\Lambda^e$ for the corresponding $e\in E(T)$.

\begin{prop}\label{prop:linear} The linear equations of the form $\bar{p}_{i_1\dots i_n}=0$ hold for any phylogenetic $n$-leaf tree $T$ evolving under a TN93 model when 
    \begin{itemize}
        \item[(i)] exactly one $i_k$ is equal to $3$ or $4$, or
        \item[(ii)] exactly one $i_k$ is equal to $2$ and the rest are equal to $1.$         
    \end{itemize} 
\end{prop}

\begin{proof}

    Case $n=3$ is proven by \Cref{lem:tripodq}. If $T$ is a tree with $n$ leaves and $n>3$, we use as induction hypothesis that the equations hold for trees with less than $n$ leaves. 
    Since $n>3$, $T$  has a cherry that does not contain the distinguished element. Let $T_1$ be the tripod formed from this cherry and consider the decomposition $T=T_1\ast T_2$, where is an $(n-1)$-leaf tree. Since there will be a single distinguished element in $T_2$, we can assume without loss of generality that it is leaf $l_n$ and that the cherry considered in $T_1$  has leaves $l_1$ and $l_2$ (reordering indices if necessary).
    
    In case (i), consider $l\in\{3,4\}$ and $i_k\neq l$ for $1\leq k\leq n-1$. By the induction hypothesis, $\bar{p}^{T_2}_{j i_{3}\dots i_{n-1}l}=0$ for $j\neq l$ and $\bar{p}^{T_1}_{i_1i_2 l}=0$. Therefore,
    \begin{equation*}
    \bar{p}_{i_1\dots i_{n-1}l}=\sum_{j\in \Sigma} \langle u^j,u^j\rangle
   \, \bar{p}^{T_1}_{i_1i_2 j}\bar{p}^{T_2}_{j i_{3}\dots i_{n-1}l}=\langle u^l,u^l\rangle\bar{p}^{T_1}_{i_1i_2 l}\bar{p}^{T_2}_{l i_{3}\dots i_{n-1}l}=0.
    \end{equation*}
    In case (ii) note that $\bar{p}^{T_i}_{1...1 j}=0$ for any $j\neq 1$ by the induction hypothesis and case (i). Hence, we get    
    $\bar{p}_{1\dots 12}=\sum_{j\in \Sigma} \langle u^j,u^j\rangle
   \, \bar{p}^{T_1}_{11 j}\bar{p}^{T_2}_{j 1\dots 1 2}=0.$
 \end{proof}
 
We pose the following question.
{\begin{question}\label{qu:ModelInvariants}
Do the equations in \Cref{prop:linear} determine a system of generators for the set of linear model invariants for $n$-leaf trees?
\end{question}}
In the case of quartets, this is proven in \Cref{cor:quartet_linear}. This problem is related to the space of phylogenetic mixtures and model selection, see \cite{CFK}, and we expect to address it {for $n$-leaf trees} in a forthcoming paper.

By decomposing a tree into a tripod and the remaining subtree as in the proof of \Cref{prop:linear}, we can obtain other linear invariants for $n\geq 4$ based on certain leaf configurations of the tree.

\begin{lemma}\label{lem:linear_top}
Let $T$ be a tree evolving under a TN93 model.  
If nodes $l_j$ and $l_k$ form a cherry, coordinates $\bar{p}_{i_1\dots i_n}$ where $\{i_j,i_k\}=\{3,4\}$ vanish for any $p\in \im \varphi_T$. 
\end{lemma}
\begin{proof}
Without loss of generality we can assume that $l_1$ and $l_2$ form a cherry as above (such that $i_{1}=3$ and $i_2=4$) and we can view $T$ as gluing a tripod tree $T_1$ and an $(n-1)$-leaf tree $T_2$.
Then
\begin{equation*}
    \bar{p}_{34i_3\dots i_{n}}=\sum_{j\in \Sigma} \langle u^j,u^j\rangle
   \, \bar{p}^{T_1}_{ 34j}\bar{p}^{T_1}_{ji_3\dots i_{n}},
    \end{equation*}
which is zero by \Cref{rem:tripod_lin_inv}.
\end{proof}

When both $3$ and $4$ appear exactly once, $\bar{p}_{i_1\dots i_{n-2}34}$ is a model invariant as proved in \Cref{prop:linear}. As we will see in \Cref{cor:quartet_linear}, if both $3$ and $4$ appear twice, $\bar{p}_{i_1\dots i_4}$ yield topology invariants for quartets. 
A natural question that remains to be addressed for larger trees is the following:
\begin{question}
Are equations $\bar{p}_{i_1\dots i_n}=0$ topology invariants for $n$-leaf trees if both $3$ and $4$ appear at least twice?
\end{question}

Next we go beyond linear invariants. Any phylogenetic invariant of the tripod can easily be extended to a model invariant for trees with $n$ leaves:

\begin{lemma}\label{lem:tripodmodel} Let 
$f(\lbrace\bar{p}_{ijk}\rbrace)$
be a phylogenetic invariant of the tripod. 
Then, for trees with $n\geq 3$ leaves evolving under TN93,
$f(\lbrace\bar{p}_{ijk1\dots1}\rbrace)$
is a model invariant.
\end{lemma}




\begin{proof}
Let $f\in \mathbb{C}[\bar{p}_{ijk}\mid i,j,k\in\Sigma]$ be a phylogenetic  invariant for the tripod and let $\tilde{f}\in\mathbb{C}[\bar{p}_{ijk1\dots 1}\mid i,j,k\in\Sigma]$ be the extension of $f$ via $\bar{p}_{ijk}\mapsto\bar{p}_{ijk1\dots 1}$.
The fact that $\tilde{f}$ is a model invariant for $n$-leaf trees follows directly from \Cref{lem:marginal}. 
Indeed, let $p$ be a tensor in $\im \psi_T$ for any tree $T$. 
 By marginalizing $p\in\otimes^n W$ over the last $n-3$ components, we have a tensor
$p^+\in\im \psi_{T_3}$ on the tripod $T_3$. 
In  coordinates in the basis $B_n$ we obtain $\tilde{f}(\lbrace\bar{p}_{ijk1\dots 1}\rbrace)=f(\lbrace\overline{p^+}_{ijk}\rbrace)$ and it vanishes because $f$ is a model invariant for tripods.
\end{proof}

\begin{exmp}\label{ex:InvariantExtension}\rm The tripod invariants in \Cref{lem:tripodp} can be extended to model invariants for quartets as follows:
\begin{gather*}
          \bar{p}_{2221}\bar{p}_{4411}-\frac{\p{34}-\p{12}}{\p{34}}\bar{p}_{2211}\bar{p}_{4421}, \quad
        \bar{p}_{2221}\bar{p}_{4141}-\frac{\p{34}-\p{12}}{\p{34}}\bar{p}_{2121}\bar{p}_{4241},\\        \bar{p}_{2221}\bar{p}_{1441}-\frac{\p{34}-\p{12}}{\p{34}}\bar{p}_{1221}\bar{p}_{2441}, \quad
\bar{p}_{3321}\bar{p}_{4411}+\frac{\p{12}}{\p{34}}\bar{p}_{3311}\bar{p}_{4421},\\
\bar{p}_{3231}\bar{p}_{4141}+\frac{\p{12}}{\p{34}}\bar{p}_{3131}\bar{p}_{4241}, \quad 
\bar{p}_{2331}\bar{p}_{1441}+\frac{\p{12}}{\p{34}}\bar{p}_{1331}\bar{p}_{2441},\\
\bar{p}_{1441}\bar{p}_{4141}\bar{p}_{441}-\frac{\p{1}\p{2}\p{12}}{(\p{1}-\p{2})^2}\bar{p}_{1111}\bar{p}_{4441}^2, \quad 
\bar{p}_{1331}\bar{p}_{3131}\bar{p}_{331}-\frac{\p{3}\p{4}\p{34}}{(\p{3}-\p{4})^2}\bar{p}_{1111}\bar{p}_{3331}^2,\\  \bar{p}_{3321}\bar{p}_{3231}\bar{p}_{2331}-\frac{\p{3}\p{4}\p{12}^2}{(\p{3}-\p{4})^2(\p{12}-\p{34})}\bar{p}_{2221}\bar{p}_{3331}^2.
\end{gather*}
\end{exmp}

\subsection{Invariants for quartets}


We now turn our attention to quartets. We call $l_il_j|l_kl_m$ the trivalent tree with four leaves $l_i,l_j,l_k,l_m$ whose interior edge separates leaves $l_i,l_j$ from $l_k,l_m$ (that is, $l_i,l_j$ and $l_k,l_m$ are cherries in this tree).  We will focus on the tree $l_1l_2|l_3l_4$ but all the results in this section are analogous for the other two tree topologies $l_1l_3|l_2l_4$ and $l_1l_4|l_2l_3$.
By $\varphi_T(\Lambda^1,\dots,\Lambda^4,\Lambda)$ we mean that $\Lambda^i$ is assigned to the edge incident to leaf $l_i$ for $i=1,\dots,4$, and $\Lambda$ is assigned to the interior edge.

\begin{prop}\label{cor:quartet_linear} The following 172 linear equations hold for any quartet $T$ evolving under the evolutionary model $\mathcal{M}=TN93$: 
\begin{itemize}
    \item[(i)] $\bar{p}_{i_1i_2i_3i_4}=0$ if exactly one $i_k$ is equal to 3 or 4 for $k\in\lbrace 1,2,3,4\rbrace$,
    \item[(ii)] $\bar{p}_{1112}=\bar{p}_{1121}=\bar{p}_{1211}=\bar{p}_{2111}=0$,
\end{itemize}
and these equations generate all linear model invariants. Moreover, if $T=l_1l_2|l_3l_4$, 
$\bar{p}_{3434},\bar{p}_{3443},\bar{p}_{4334}$, and $\bar{p}_{4343}$
are linear topology invariants of $T$.
\end{prop}
\begin{proof}
 As a a particular case of \Cref{prop:linear} we obtain the list of model invariants in (i) and (ii).  On the other hand, from Lemma \ref{lem:linear_top} 
we get that $\bar{p}_{3434},\bar{p}_{3443},\bar{p}_{4334}$, and $\bar{p}_{4343}$ vanish when $p$ evolves on $T={l_1l_2|l_3l_4}$.  It can be easily checked that $\bar{p}_{3434},\bar{p}_{3443},\bar{p}_{4334}$, and $\bar{p}_{4343}$ are generically non-zero for either tree ${l_1l_3|l_2l_4}$ or ${l_1l_4|l_2l_3}$, and hence they are topology invariants for $T$.
 For example, if $T=l_1l_3|l_2l_4$ let $q=\varphi_T(Id,Id,Id,Id,\Lambda)$ for a diagonal matrix $\Lambda$,  then
\begin{equation}\label{eq:3434}
\bar{q}_{3434}=\bar{q}_{4343}=\frac{\pi_{12}\pi_{34}}{\pi_1\pi_2\pi_3\pi_4}(\lambda_1 -\lambda_2)
\end{equation}
(see Lemma \ref{lem:almost_monomial} in  \Cref{sec:app_quartets}), 
which is non-zero if $\lambda_1\neq \lambda_2.$

Let $\mathcal{L}_4$ be the linear space defined by the 172 equations in (i) and (ii). As these are linearly independent equations (they involve different coordinates), $\mathcal{L}_4$ has dimension $256-172=84$. In what follows we prove that there is a subset of points in $CV_{l_1l_2|l_3l_4}\cup CV_{l_1l_3|l_2l_4}\cup CV_{l_1l_4|l_2l_3}$ that spans $\mathcal{L}_4$. 

For each $j\in \Sigma=\{1,2,3,4\}$, set $D_j=\mathrm{diag}(e^j)$.
Consider the $84$ 4-tuples  not appearing as a subindex in the equations of $\mathcal{L}_4$ and let $\mathbf{i}=(i_1,\dots,i_4)$ be any of these 4-tuples. 

If $\mathbf{i}$ is different from $(3,4,3,4),(4,3,4,3), (3,4,4,3), (4,3,3,4)$, take $T=l_1l_2|l_3l_4$ and a diagonal matrix $\Lambda$, and 
 define the point 
 \[{p}^{\,\mathbf{i}}={\varphi}_T(D_{i_1},D_{i_2},D_{i_3},D_{i_4},\Lambda).\]
By \eqref{eq:id}, the unique possibly non-zero coordinate of ${p}^{\,\mathbf{i}}$ in the basis $B_4$ is $i_1\dots i_4$. Using the expressions given in  \Cref{sec:app_quartets}, we see that this coordinate is non-zero if $\Lambda$ has non-zero generic elements in the diagonal. 
Therefore, we have exactly 80 linearly independent points in this set.

If $\mathbf{i}=(3,4,3,4)$ or $(4,3,4,3)$ consider $T=l_1l_3|l_2l_4$ and the point 
${p}^{\,\mathbf{i}}=\varphi_T(D_{i_1},\allowbreak  D_{i_2}, \allowbreak D_{i_3}, D_{i_4}, \Lambda)$, where $\Lambda=\diag(\lambda_1,\lambda_2,\lambda_3,\lambda_4)$ has $\lambda_2\neq \lambda_1$. 
Note these are two linearly independent points whose $B_4$ coordinates are all zero except those indexed by $3434$, $4343$, respectively, which coincide with expression \eqref{eq:3434}.

If $\mathbf{i}=(3,4,4,3)$ or $(4,3,3,4)$ an analogous argument applies by considering points ${p}^{\,\mathbf{i}}$ in the variety of $T=l_1l_4|l_2l_3$.
 
The 80 linearly independent points above together with the four points ${p}^{\,(3,4,3,4)}$, ${p}^{\,(4,3,4,3)}$, ${p}^{\,(3,4,4,3)}$, and ${p}^{\,(4,3,3,4)}$ form a set of 84 linearly independent points in $\mathcal{L}_4$ because all have a single-non zero coordinate and all of them are in different positions.
\end{proof}

\begin{remark}\label{rm:linearspan}\rm 
From the previous result we get that the 84-dimensional space $\mathcal{L}_4$ defined as the set of tensors where the 172 equations in (i) and (ii) vanish coincides with the linear span of $CV_{l_1l_2|l_3l_4}\cup CV_{l_1l_3|l_2l_4}\cup CV_{l_1l_4|l_2l_3}.$ If we add equations  $\bar{p}_{3434}=0,\bar{p}_{3443}=0,\bar{p}_{4334}=0$, and $\bar{p}_{4343}=0$. then the zero set $\mathcal{L}_{l_1l_2|l_3l_4}$ of dimension 80 is the linear span of $CV_{l_1l_2|l_3l_4}.$
The table below displays which equations hold for each of the three possible quartet topologies, thus providing their topology invariants.

$$\begin{array}{||c|c|c|c|c|c|c||}
    \hline
     &  \bar{p}_{3344}=0 &\bar{p}_{4433}=0 & \bar{p}_{3434}=0 & \bar{p}_{4343}=0 &  \bar{p}_{3443}=0 & \bar{p}_{4334}=0 \\ 
     \hline
     \hline
     l_1l_2|l_3l_4 & No & No & Yes & Yes & Yes & Yes \\
     l_1l_3|l_2l_4 & Yes & Yes & No & No & Yes & Yes \\
     l_1l_4|l_2l_3 &  Yes & Yes & Yes & Yes & No & No \\
     \hline
\end{array}$$


\end{remark}

\begin{lemma}\label{lem:rank} Let $T=l_1l_2|l_3l_4$ be a quartet and consider the flattening matrix $\flatt(\bar{p})$. Then, for any $i,j\in \Sigma$, column $(i,j)$ is a linear combination of columns $(1,1),(1,2),(1,3),$ and $(1,4)$. Moreover, for generic $\pi$, there is an open set $\mathcal{U}\subseteq CV_T$ containing the no-evolution point $p^0$ such that, for any $p\in \mathcal{U}$, $\mathrm{rank}\left(\flatt(\bar{p})\right)=4$ and
\begin{enumerate}[(i)]
    \item the submatrix formed by columns $(1,j),(j,1),(2,j),(j,2)$ has rank 1 for $j\in\{3,4\}$,
    \item the submatrix formed by columns $(1,2),(2,1)$ has rank 1,
    \item the submatrix formed by columns $(1,1),(1,2),(2,2)$ has rank 2,
    \item the submatrix formed by columns $(1,1),(1,2),(1,j),(j,j)$ has rank 3 for $j\in\{3,4\}$.
\end{enumerate}
\end{lemma}
The submatrices described in the statement 
 are displayed in Tables \ref{tab:quadrics} to  \ref{tab:quarticsT} in \Cref{app:flattening}. Note that the generic rank four claimed here was already known by a more detailed proof of Theorem \ref{thm:ARflat} given in \cite{Snyman}.
 
\begin{proof} 
Given $\bar{p}=\varphi_T(\Lambda^1,\dots, \Lambda^4,\Lambda)$, consider $\bar{q}=\varphi_T(Id,\dots,Id, \Lambda)$ so that 
\[\bar{p}_{i_1 i_2 i_3 i_4}=\lambda^1_{i_1}\lambda^2_{i_2}\lambda^3_{i_3}\lambda^4_{i_4}\bar{q}_{i_1i_2i_3 i_4}\] for any $i_j \in \Sigma.$ 
In standard coordinates $\bar{q}$ is $q=\psi_T(Id,\dots,Id, A^{-t}\Lambda A^t)$, which can be written as the gluing $q^{T_1}\ast q^{T_2}$, where $q^{T_1}$ evolves on the tripod $T_1$ with the identity matrix at leaves 1, 2 and matrix $A^{-t}\Lambda A^t$ at the third leaf, and $q^{T_2}$ is the no-evolution point on the tripod $T_2$.
Note that $q^{T_1}$ coincides with the marginalization $q^+$ of $q$ over leaf $l_3$. By \Cref{lem:marginal}, we have $\bar{q}^{T_1}_{i_1i_2k}=\overline{q^+}_{i_1i_2k}=\bar{q}_{i_1i_21k}.$ Therefore using Theorem \ref{thm:gluing} we have
\[\bar{q}_{i_1i_2 ij}=\sum_{k\in\Sigma} \langle u^k,u^k\rangle
   \, \bar{q}^{T_1}_{i_1i_2k}\bar{q}^{T_2}_{kij}   
   =\sum_{k\in\Sigma} \langle u^k,u^k\rangle
   \, \bar{q}^{T_2}_{kij} \bar{q}_{i_1 i_2 1 k}\]
for any $i_1,i_2,i,j \in \Sigma$. In particular,
column $(i,j)$ of $\flatt(\bar{q})$ is a linear combination of columns $(1,k)$ for $k\in\Sigma$.
In \Cref{table:flatq} we display $\flat_{l_1l_2|l_3l_4}(\bar{q})$ to visualize the submatrices in the statement. 



Now we have 
\begin{multline}\label{eq:pbar}
\bar{p}_{i_1i_2 i j}=\lambda^1_{i_1}\lambda^2_{i_2}\lambda^3_{i}\lambda^4_{j}\bar{q}_{i_1i_2 i j}=\sum_{k\in\Sigma} \langle u^k,u^k\rangle
   \, \lambda^3_{i}\lambda^4_{j}\bar{q}^{T_2}_{kij} \lambda^1_{i_1}\lambda^2_{i_2}\bar{q}_{i_1 i_2 1 k}\\
   =\sum_{k\in\Sigma}\langle u^k,u^k\rangle
   \, \lambda^3_{i}\lambda^4_{j}\bar{q}^{T_2}_{kij} (\lambda^3_1\lambda^4_{k})^{-1}\bar{p}_{i_1 i_2 1 k},
\end{multline}
\noindent where the last equality holds if $\lambda^3_1\neq 0$ and $\Lambda^4$ is invertible.
Thus, on an open set of $V_T$ (and hence on the whole variety), column $(i,j)$ of $\flatt(\bar{p})$ is a linear combination of
columns $(1,1)$, $(1,2)$, $(1,3)$, and $(1,4)$. 
Moreover, the rank of $\flatt(\bar{p})$ is exactly 4 in an open set containing the no-evolution point, since the 4-minor formed by rows and columns $(1,k)$ does not vanish at $p^0$.

The non-vanishing coordinates of the no-evolution point $q^{T_2}$ in \Cref{lem:tripodq} determine how many non-vanishing summands are there in \eqref{eq:pbar}, and hence yield the rank of submatrices (i)-(iv).
\end{proof}

For the TN93 process on the quartet $T=l_1l_2|l_3l_4$ we have 5 transition matrices with 3 free parameters each, hence $\dim V_T=15$ and $\dim CV_T=16$ (see Subsection \ref{sec:alg_var}). By \Cref{rm:linearspan}, $CV_T$ lies in the linear space $\mathcal{L}_{l_1l_2|l_3l_4}$ of dimension $80$. Next we provide $\textrm{codim}(CV_T)=64$ elements in the ideal of $CV_T$ that define the variety locally at the no-evolution point. 

Inspired by the results in \cite[Theorem 5.4]{CFM} that provide local equations for equivariant models, we consider invariants arising from 
\begin{itemize}[leftmargin=20pt]
    \item[(a)] extending the tripod equations in \Cref{lem:tripodp} by adding $1$ in the first and third position respectively (which are phylogenetic invariants for $T$ by \Cref{lem:tripodmodel}), and
    \item[(b)] rank constraints on  $\flatt(\bar{p})$ as in \Cref{lem:rank}; more precisely, for each of the submatrices of rank $r$, $r=1,2,3$,  in \Cref{lem:rank}, consider a non-vanishing $r$-minor (namely one containing only rows and columns of type $(1,j)$ for $j\in\Sigma$), and consider all $(r+1)$-minors containing it (check Tables \ref{tab:quadrics} to  \ref{tab:quarticsT} in  \Cref{app:flattening} 
for the precise description of these minors).
\end{itemize}
This gives the 64 phylogenetic invariants listed in the statement below. 
\begin{thm}\label{lem:localCIquartet} Consider the tree $T={l_1l_2\mid l_3l_4}$ and let it evolve under the TN93 model. Then there exist 64 equations that cut out the variety $CV_T$ on an open set containing the no-evolution point $p^0$, arising from 

\begin{itemize}[leftmargin=10pt]
\item the extension of the 6 quadrics and 3 cubics in \cref{lem:tripodp} with $1$ in the first leaf,
\item the extension of the 6 quadrics and 3 cubics in \cref{lem:tripodp} with $1$ in the third leaf, 
\item 2-minors of columns $(1,j),(j,1),(2,j),(4,j)$ of $\allowbreak\flatt(\bar{p})$ for each $j\in\{3,4\}$ (12 quadrics),
\item 2-minors of columns $(1,2),(2,1)$ of  $\flatt(\bar{p})$ (4 quadrics),
\item 3-minors of columns $(1,1),(1,2),(2,2)$ of  $\flatt(\bar{p})$ (4 cubics),
\item 4-minors of columns $(1,1),(1,2),(1,j),(j,j)$ of  $\flatt(\bar{p})$, for each $j\in\{3,4\}$ (7 quartics).
\end{itemize}
\end{thm}

\begin{proof}


\texttt{Macaulay2} computations show that the Jacobian of these polynomials at $p^0$ is indeed 64. Thus, on a Zariski open subset containing $p^0$, these polynomials define a complete intersection of dimension 16 that coincides with $CV_T$.
\end{proof}

\section{Discussion}\label{sec:discussion}

We have introduced a new approach to work with algebraic time-reversible models that have a given stationary distribution $\pi$. We assume that this $\pi$ can be inferred from data, that is, the given data has reached the equilibrium distribution. We also assumed that this stationary distribution was the same as the one that initiated the process (as it is usual to assume on a time-reversible process). This is an important assumption for our methods to work: if the distribution at the root $\pi^r$ was supposed to be different from $\pi$, the statements of our main results would not hold. It would be interesting to explore a new model that would allow $\pi^r$ to be parameters as well. 

We have illustrated our methods with an insight into the TN93 model. Far from providing an extensive work on this model, we have mainly worked on tripods and quartet trees. We are aware that the tools presented here can allow the extension of this work to trees on any number of leaves and we aim to develop such work in a forthcoming project. Another project is exploring the tools we have developed with a view towards model selection by using linear model invariants of different ATR models and studying the space of phylogenetic mixtures (in the sense of \cite{KDGC} and \cite{CFK}).

\bibliographystyle{alpha}
\bibliography{biblio_ATR}

\appendix

\section{Computations}\label{app:Computations}

\subsection{Gluing parameters}\label{app:gluing}
\begin{proof}[Proof of~\Cref{lem:Ys}] By rooting the tree  $T'$ at $s$ we have
   \[p^T_{\bm{i}_{\alpha},k,\bm{i}_{\beta}}=\sum_{\bm{j} \in ext(\bm{i}_{\alpha},k,\bm{i}_{\beta})} \pi_{k}\prod_{e\in E(T')} M^e_{j_{p(e)},j_{c(e)}} =\]
   \[=\frac{1}{\pi_{k}}\sum_{ \begin{array}{c} \scriptstyle v \in Int(T_1)\cup Int(T_2) \\
\scriptstyle j_v \in \Sigma\end{array}}\left(\pi_{k}\prod_{e\in E(T_1)} M^e_{j_{p(e)},j_{c(e)}}\right)\left(\pi_{k}\prod_{e\in E(T_2)} M^e_{j_{p(e)},j_{c(e)}}\right)=\]
   \[=\frac{1}{\pi_{k}}\left(\sum_{ \begin{array}{c} \scriptstyle v \in Int(T_1) \\
\scriptstyle j_v \in \Sigma\end{array}}\pi_{k}\prod_{e\in E(T_1)} M^e_{j_{p(e)},j_{c(e)}}\right)\left(\sum_{ \begin{array}{c} \scriptstyle v \in Int(T_2) \\
\scriptstyle j_v \in \Sigma\end{array}}\pi_{k}\prod_{e\in E(T_2)} M^e_{j_{p(e)},j_{c(e)}}\right)\]
   and the claim follows.
\end{proof}

\subsection{Computations for the tripod}

The list of 9 quadric binomials, 29 cubic binomials, and 3 quintic binomials generating $I_T$ for a tripod $T$ mentioned in~\Cref{lem:tripodp} is:

\vspace{0.3cm}
\noindent \textbf{Generators of degree 2}
    \begin{multicols}{2}
        \begin{enumerate}[leftmargin=*,label=\textbf{\small (\arabic*)}]
        \setlength\itemsep{0.4em}
            \item $\bar{p}_{222}\bar{p}_{441}-\frac{\pi_{34}-\pi_{12}}{\pi_{34}}\Bar{p}_{221}\Bar{p}_{442}$
            \item $\bar{p}_{332}\bar{p}_{441}+\frac{\pi_{12}}{\pi_{34}}\bar{p}_{331}\bar{p}_{442}$
            \item $\bar{p}_{222}\bar{p}_{414}-\frac{\pi_{34}-\pi_{12}}{\pi_{34}}\bar{p}_{212}\bar{p}_{424}$
            \item $\bar{p}_{323}\bar{p}_{414}+\frac{\pi_{12}}{\pi_{34}}\bar{p}_{313}p_{424}$
            \item $\bar{p}_{144}\bar{p}_{222}-\frac{\pi_{34}-\pi_{12}}{\pi_{34}} \bar{p}_{122}\bar{p}_{244}$
            \item $\bar{p}_{332}\bar{p}_{221}-\frac{\pi_{12}}{\pi_{12}-\pi_{34}}\bar{p}_{331}\bar{p}_{222}$
            \item $\bar{p}_{144}\bar{p}_{233}+\frac{\pi_{12}}{\pi_{34}}\bar{p}_{133}\bar{p}_{244}$
            \item $\bar{p}_{122}\bar{p}_{233}-\frac{\pi_{12}}{\pi_{12}-\pi_{34}}\bar{p}_{133}\bar{p}_{222}$
            \item $\bar{p}_{323}\bar{p}_{212}-\frac{\pi_{12}}{\pi_{12}-\pi_{34}}\bar{p}_{313}\bar{p}_{222}$
        \end{enumerate}
    \end{multicols}
\newpage
\noindent \textbf{Generators of degree 3}
    \begin{multicols}{2}
        \begin{enumerate}[leftmargin=*,label=\textbf{\small(\arabic*)}]
        \addtocounter{enumi}{9}
        \setlength\itemsep{0.4em}
            \item $\bar{p}_{144}\bar{p}_{424}\bar{p}_{442}- \frac{\pi_1\pi_2\pi_{34}}{(\pi_1-\pi_2)^2}\bar{p}_{122}\bar{p}
            _{444}^2$
            \item $\bar{p}_{244}\bar{p}_{414}\bar{p}_{442}-\frac{\pi_1\pi_2\pi_{34}}{(\pi_1-\pi_2)^2}\bar{p}_{212}\bar{p}_{444}^2$
            \item $\bar{p}_{244}\bar{p}_{424}\bar{p}_{441}- \frac{\pi_1\pi_2\pi_{34}}{(\pi_1-\pi_2)^2}\bar{p}_{221}\bar{p}_{444}^2$
            \item $\bar{p}_{144}\bar{p}_{414}\bar{p}_{441}-\frac{\pi_1\pi_2\pi_{12}}{(\pi_1-\pi_2)^2}\bar{p}_{111}\bar{p}_{444}^2$
            \item $\bar{p}_{122}\bar{p}_{414}\bar{p}_{441}-\frac{\pi_{12}}{\pi_{34}}\bar{p}_{111}\bar{p}_{424}\bar{p}_{442}$
            \item $\bar{p}_{144}\bar{p}_{212}\bar{p}_{441}- \frac{\pi_{12}}{\pi_{34}}\bar{p}_{111}\bar{p}_{244}\bar{p}_{442}$
            \item $\bar{p}_{122}\bar{p}_{212}\bar{p}_{441}-\frac{\pi_{12}}{\pi_{34}-\pi_{12}}\bar{p}_{111}\bar{p}_{222}\bar{p}_{442}$
            \item $\bar{p}_{133}\bar{p}_{212}\bar{p}_{441}+\bar{p}_{111}\bar{p}_{233}\bar{p}_{442}$
            \item $\bar{p}_{122}\bar{p}_{313}\bar{p}_{441}+\bar{p}_{111}\bar{p}_{323}\bar{p}_{442}$
            \item $\bar{p}_{122}\bar{p}_{313}\bar{p}_{441}+\bar{p}_{111}\bar{p}_{323}\bar{p}_{442}$
            \item $\bar{p}_{144}\bar{p}_{221}\bar{p}_{414}-\frac{\pi_{12}}{\pi_{34}}\bar{p}_{111}\bar{p}_{244}\bar{p}_{424}$
            \item $\bar{p}_{122}\bar{p}_{221}\bar{p}_{414}-\frac{\pi_{12}}{\pi_{34}-\pi_{12}}\bar{p}_{111}\bar{p}_{222}\bar{p}_{424}$
            \item $\bar{p}_{133}\bar{p}_{221}\bar{p}_{414}+\bar{p}_{111}\bar{p}_{233}\bar{p}_{424}$
            \item $\bar{p}_{122}\bar{p}_{331}\bar{p}_{414}+\bar{p}_{111}\bar{p}_{332}\bar{p}_{424}$
            \item $\bar{p}_{144}\bar{p}_{212}\bar{p}_{221}-\frac{\pi_{12}}{\pi_{34}-\pi_{12}}\bar{p}_{111}\bar{p}_{222}\bar{p}_{244}$
            \item $\bar{p}_{122}\bar{p}_{212}\bar{p}_{221}-\frac{\pi_{12}\pi_{34}}{(\pi_{12}-\pi_{34})^2}\bar{p}_{111}\bar{p}_{222}^2$
            \item $\bar{p}_{133}\bar{p}_{212}\bar{p}_{221}-\frac{\pi_{34}}{\pi_{12}-\pi_{34}}\bar{p}_{111}\bar{p}_{233}\bar{p}_{222}$
            \item $\bar{p}_{144}\bar{p}_{313}\bar{p}_{221}+\bar{p}_{111}\bar{p}_{323}\bar{p}_{244}$
            \item $\bar{p}_{122}\bar{p}_{313}\bar{p}_{221}-\frac{\pi_{34}}{\pi_{12}-\pi_{34}}\bar{p}_{111}\bar{p}_{323}\bar{p}_{222}$
            \item $\bar{p}_{331}\bar{p}_{323}\bar{p}_{233}-\frac{\pi_3\pi_4\pi_{12}}{(\pi_3-\pi_4)^2}\bar{p}_{333}^2\bar{p}_{221}$
            \item $\bar{p}_{111}\bar{p}_{323}\bar{p}_{233}-\frac{\pi_{12}}{\pi_{34}}\bar{p}_{133}\bar{p}_{313}\bar{p}_{221}$
            \item $\bar{p}_{144}\bar{p}_{331}\bar{p}_{212}+\bar{p}_{111}\bar{p}_{332}\bar{p}_{244}$
            \item $\bar{p}_{122}\bar{p}_{331}\bar{p}_{212}-\frac{\pi_{34}}{\pi_{12}-\pi_{34}}\bar{p}_{111}\bar{p}_{332}\bar{p}_{222}$
            \item $\bar{p}_{133}\bar{p}_{331}\bar{p}_{212}-\frac{\pi_{34}}{\pi_{12}}\bar{p}_{111}\bar{p}_{332}\bar{p}_{233}$
            \item $\bar{p}_{122}\bar{p}_{333}^2-\frac{(\pi_3-\pi_4)^2}{\pi_3\pi_4\pi_{12}}\bar{p}_{133}\bar{p}_{332}\bar{p}_{323}$
            \item $\bar{p}_{122}\bar{p}_{313}\bar{p}_{331}-\frac{\pi_{34}}{\pi_{12}}\bar{p}_{111}\bar{p}_{332}\bar{p}_{323}$
            \item $\bar{p}_{133}\bar{p}_{313}\bar{p}_{331}-\frac{\pi_3\pi_4\pi_{34}}{(\pi_3-\pi_4)^2}\bar{p}_{111}\bar{p}_{333}^2$
        \end{enumerate}
    \end{multicols}
    \begin{enumerate}[leftmargin=*,label=\textbf{\small(\arabic*)}]
        \addtocounter{enumi}{36}
        \setlength\itemsep{0.5em}
        \item $\bar{p}_{244}\bar{p}_{424}\bar{p}_{442}-\frac{\pi_1\pi_2(\pi_3+\pi_4)^2}{(\pi_1-\pi_2)^2(\pi_{34}-\pi_{12})} \bar{p}_{222}\bar{p}_{444}^2$
        \item $\bar{p}_{332}\bar{p}_{323}\bar{p}_{233}-\frac{\pi_3\pi_4\pi_{12}^2}{(\pi_3-\pi_4)^2(\pi_{12}^2-\pi_{34}^4)}\bar{p}_{333}^2\bar{p}_{222}$
        \item $\bar{p}_{333}^2\bar{p}_{212}-\frac{(\pi_{12}-\pi_{34})(\pi_3-\pi_4)^2}{\pi_3\pi_4\pi_{12}}\bar{p}_{313}\bar{p}_{332}\bar{p}_{233}$
    \end{enumerate}
\vspace{0.3cm}
\noindent \textbf{Generators of degree 5}

    \begin{enumerate}[leftmargin=*,label=\textbf{\small (\arabic*)}]
    \addtocounter{enumi}{39}
    \setlength\itemsep{0.5em}
        \item $\bar{p}_{144}\bar{p}_{333}^2\bar{p}_{414}\bar{p}_{442}+\frac{\pi_1\pi_2(\pi_3-\pi_4)^2}{\pi_3\pi_4(\pi_1-\pi_2)^2}\bar{p}_{133}\bar{p}_{313}\bar{p}_{332}\bar{p}_{444}^2$
        \item $\bar{p}_{144}\bar{p}_{333}^2\bar{p}_{424}\bar{p}_{441}+\frac{\pi_1\pi_2(\pi_3-\pi_4)^2}{\pi_3\pi_4(\pi_1-\pi_2)^2}\bar{p}_{133}\bar{p}_{331}\bar{p}_{323}\bar{p}_{444}^2$
        \item $ \bar{p}_{333}^2\bar{p}_{244}\bar{p}_{414}\bar{p}_{441}+\frac{\pi_1\pi_2(\pi_3-\pi_4)^2}{\pi_3\pi_4(\pi_1-\pi_2)^2}\bar{p}_{313}\bar{p}_{331}\bar{p}_{233}\bar{p}_{444}^2 $
    \end{enumerate}


\subsection{Computations for quartets}\label{sec:app_quartets}

In the following lemma we compute the coordinates in $B_n$ for any point in evolving on $T=12|34$ under the TN93 model. The coordinates for points on the variety of $T=13|24$ or $14|23$ can be obtained by correspondingly permuting the subindices of the coordinates. This result is used in  Proposition \Cref{cor:quartet_linear}. 
\begin{lemma}\label{lem:almost_monomial} Consider the tree $T=12|34$ evolving under the evolutionary model $\mathcal{M}=TN93$. In the basis $B_4$, any tensor $p=\varphi_T(\Lambda^1,\dots,\Lambda^5)$ has monomial coordinates except for $\bar{p}_{iiii}$ 
 and $\bar{p}_{iijj}$ for $i,j\in\{2,3,4\}$. These coordinates can be obtained by \eqref{eq:id} from the following expressions of the coordinates of $q=\varphi_T(Id,Id,Id,Id,\Lambda)$, where $\Lambda=\diag(\lambda_1,\lambda_2,\lambda_3,\lambda_4)$ (non listed coordinates are zero):
 \begin{enumerate}[label=\textbf{\small (\arabic*)}]
 \setlength\itemsep{0.5em}
    \item $\bar{q}_{1111}=\lambda_1$
    \item $\bar{q}_{1122}=\bar{q}_{2211}=\frac{1}{\pi_{12}\pi_{34}}\lambda_1$
    \item $\bar{q}_{1133}=\bar{q}_{3311}=\frac{\pi_{34}}{\pi_3\pi_4}\lambda_1$
    \item $\bar{q}_{1144}=\bar{q}_{4411}=\frac{\pi_{12}}{\pi_1\pi_2}\lambda_1$
    \item $\bar{q}_{1212}=\bar{q}_{1221}=\bar{q}_{2112}=\bar{q}_{2121}=\frac{1}{\pi_{12}\pi_{34}}\lambda_2$
    \item $\bar{q}_{1222}=\bar{q}_{2122}=\bar{q}_{2212}=\bar{q}_{2221}=\frac{\pi_{34}-\pi_{12}}{\pi_{34}^2\pi_{12}^2}\lambda_2$
    \item $\bar{q}_{1233}=\bar{q}_{2133}=\bar{q}_{3312}=\bar{q}_{3321}=-\frac{1}{\pi_3\pi_4}\lambda_2$
    \item $\bar{q}_{1244}=\bar{q}_{2144}=\bar{q}_{4412}=\bar{q}_{4421}=\frac{1}{\pi_1\pi_2}\lambda_2$
    \item $\bar{q}_{1313}=\bar{q}_{1331}=\bar{q}_{3113}=\bar{q}_{3131}=\frac{\pi_{34}}{\pi_3\pi_4}\lambda_3$
    \item $\bar{q}_{1323}=\bar{q}_{1332}=\bar{q}_{2313}=\bar{q}_{2331}=\bar{q}_{3123}=\bar{q}_{3132}=\bar{q}_{3213}=\bar{q}_{3231}=-\frac{1}{\pi_3\pi_4}\lambda_3$
    \item $\bar{q}_{1333}=\bar{q}_{3133}=\bar{q}_{3313}=\bar{q}_{3331}=\frac{\pi_{34}(\pi_4-\pi_3)}{\pi_3^2\pi_4^2}\lambda_3$
    \item $\bar{q}_{1414}=\bar{q}_{1441}=\bar{q}_{4114}=\bar{q}_{4141}=\frac{\pi_{12}}{\pi_1\pi_2}\lambda_4$
    \item $\bar{q}_{1424}=\bar{q}_{1442}=\bar{q}_{2414}=\bar{q}_{2441}=\bar{q}_{4124}=\bar{q}_{4142}=\bar{q}_{4214}=\bar{q}_{4241}=\frac{1}{\pi_1\pi_2}\lambda_4$
    \item $\bar{q}_{1444}=\bar{q}_{4144}=\bar{q}_{4414}=\bar{q}_{4441}=\frac{\pi_{12}(\pi_2-\pi_1)}{\pi_1^2\pi_2^2}\lambda_4$
    \item $\bar{q}_{2222}=\frac{1}{\pi_{12}^2\pi_{34}^2}\lambda_1 + \frac{(\pi_{12}-\pi_{34})^2}{\pi_{12}^3\pi_{34}^3}\lambda_2$  
    \item $\bar{q}_{2233}=\bar{q}_{3322}=\frac{1}{\pi_{12}\pi_3\pi_4}\lambda_1+\frac{\pi_{12}-\pi_{34}}{\pi_3\pi_4\pi_{12}\pi_{34}}\lambda_2$
    \item $\bar{q}_{2244}=\bar{q}_{4422}= \frac{1}{\pi_1\pi_2\pi_{34}}\lambda_1+\frac{\pi_{34}-\pi_{12}}{\pi_1\pi_2\pi_{12}\pi_{34}}\lambda_2$
    \item $\bar{q}_{2323}=\bar{q}_{2332}=\bar{q}_{3223}=\bar{q}_{3232}=\frac{1}{\pi_3\pi_4\pi_{34}}\lambda_3$
    \item $\bar{q}_{2333}=\bar{q}_{3233}=\bar{q}_{3323}=\bar{q}_{3332}=\frac{\pi_3-\pi_4}{\pi_3^2\pi_4^2}\lambda_3$
    \item $\bar{q}_{2424}=\bar{q}_{2442}=\bar{q}_{4224}=\bar{q}_{4242}=\frac{1}{\pi_1\pi_2\pi_{12}}\lambda_4$
    \item $\bar{q}_{2444}=\bar{q}_{4244}=\bar{q}_{4424}=\bar{q}_{4442}=\frac{\pi_2-\pi_1}{\pi_1^2\pi_2^2}\lambda_4$
    \item $\bar{q}_{3333}=\frac{\pi_{34}^2}{\pi_3^2\pi_4^2}\lambda_1 + \frac{\pi_{12}\pi_{34}}{\pi_3^2\pi_4^2}\lambda_2 + \frac{\pi_{34}(\pi_3-\pi_4)^2}{\pi_3^3\pi_4^3}\lambda_3$
    \item $\bar{q}_{3344}=\bar{q}_{4433}=\frac{\pi_{12}\pi_{34}}{\pi_1\pi_2\pi_3\pi_4}\lambda_1 - \frac{\pi_{12}\pi_{34}}{\pi_1\pi_2\pi_3\pi_4}\lambda_2$
    \item $\bar{q}_{4444}=\frac{\pi_{12}^2}{\pi_1^2\pi_2^2}\lambda_1+\frac{\pi_{12}\pi_{34}}{\pi_1^2\pi_2^2}\lambda_2+\frac{\pi_{12}(\pi_1-\pi_2)^2}{\pi_1^3\pi_2^3}\lambda_4$
\end{enumerate}
\end{lemma}

\begin{proof}
Let $T_1$ be the tripod with leaves $l_1,l_2,l_s$ and $T_2$ be the tripod with leaves $l_s,l_3,l_4$ so that $T$ is the gluing $T_1\ast T_2$. Define tensors ${q}^{T_1}=\varphi_{T_1}(Id,Id,\Lambda)$ and $\bar{p}^{T_2}=\varphi_{T_2}(Id,Id,Id)$ evolving on $T_1$ and $T_2$ respectively; hence $q=q_1*q_2$. 

By \Cref{thm:gluing} the coordinates of $q$ in the basis $B_n$ can be obtained by the matrix product 
{\tiny
\begin{equation*}
\left(\begin{array}{p{0.25cm}|cccc}
& 1 & 2 & 3 & 4\\
\hline
11 & \ast & 0 & 0 & 0\\
22 & \ast & \ast & 0 & 0\\
33 & \ast & \ast & \ast & 0\\
44 & \ast & \ast & 0 & \ast\\
12 & 0 & \ast & 0 & 0\\
21 & 0 & \ast & 0 & 0\\
13 & 0 & 0 & \ast & 0\\
31 & 0 & 0 & \ast & 0\\
23 & 0 & 0 & \ast & 0\\
32 & 0 & 0 & \ast & 0\\
14 & 0 & 0 & 0 & \ast\\
41 & 0 & 0 & 0 & \ast\\
24 & 0 & 0 & 0 & \ast\\
42 & 0 & 0 & 0 & \ast\\
34 & 0 & 0 & 0 & 0\\
43 & 0 & 0 & 0 & 0\\
\end{array}\right)
\left(\begin{array}{p{0.15cm}|cccccccccccccccc}
   & 11 & 22 & 33 & 44 & 12 & 21 & 13 & 31 & 23 & 32 & 14 & 41 & 24 & 42 & 34 & 43 \\
\hline
1 & \ast & \ast & \ast & \ast & 0 & 0 & 0 & 0 & 0 & 0 & 0 & 0 & 0 & 0 & 0 & 0\\
2 & 0 & \ast & \ast & \ast & \ast & \ast & 0 & 0 & 0 & 0 & 0 & 0 & 0 & 0 & 0 & 0\\
3 & 0 & 0 & \ast & 0 & 0 & 0 & \ast & \ast & \ast & \ast & 0 & 0 & 0 & 0 & 0 & 0\\
4 & 0 & 0 & 0 & \ast & 0 & 0 & 0 & 0 & 0 & 0 & \ast & \ast & \ast & \ast & 0 & 0
\end{array}\right)
\end{equation*} }
\noindent
in the sense that $q_{i_1i_2i_3i_4}$ is the product of row $(i_1,i_2)$ of the first matrix and column $(i_3,i_4)$ of the second matrix up to multiplication by a scalar product. 
As $*$ entries are monomials in (actually they are linear entries in some $\lambda_k$ in the first matrix and scalar entries in the second), non-monomial entries only appear when multiplying rows and columns with indices $(2,2),(3,3),(4,4)$. From this matrix product we get the entries that appear in the list.
\end{proof}

\section{Flattening matrices}\label{app:flattening}

\begin{landscape}
\begin{table}
$$\begin{array}{c|c|cccc|c|c|cc|c|cccc|cc}
& (1,1) & (1,4) & (4,1) & (2,4) & (4,2) & (4,4) & (2,2) & (1,2) & (2,1) & (3,3) & (1,3) & (3,1) & (2,3) & (3,2) & (3,4) & (4,3)\\
\hline
(1,1) & \bar{q}_{1111} & 0 & 0 & 0 & 0 & \bar{q}_{1144} & \bar{q}_{1122} & 0 & 0 & \bar{q}_{1133} & 0 & 0 & 0 & 0 & 0 & 0\\
\hline
(1,4) & 0 & \bar{q}_{1414} & \bar{q}_{1414} & \bar{q}_{1424} & \bar{q}_{1424} & \bar{q}_{1444} & 0 & 0 & 0 & 0 & 0 & 0 & 0 & 0 & 0 & 0\\
(4,1) & 0 & \bar{q}_{1414} & \bar{q}_{1414} & \bar{q}_{1424} & \bar{q}_{1424} & \bar{q}_{1444} & 0 & 0 & 0 & 0 & 0 & 0 & 0 & 0 & 0 & 0\\
(2,4) & 0 & \bar{q}_{1424} & \bar{q}_{1424} & \bar{q}_{2424} & \bar{q}_{2424} & \bar{q}_{2444} & 0 & 0 & 0 & 0 & 0 & 0 & 0 & 0 & 0 & 0\\
(4,2) & 0 & \bar{q}_{1424} & \bar{q}_{1424} & \bar{q}_{2424} & \bar{q}_{2424} & \bar{q}_{2444} & 0 & 0 & 0 & 0 & 0 & 0 & 0 & 0 & 0 & 0\\
\hline
(4,4) & \bar{q}_{1144} & \bar{q}_{1444} & \bar{q}_{1444} & \bar{q}_{2444} & \bar{q}_{2444} & \bar{q}_{4444} & \bar{q}_{2244} & \bar{q}_{1244} & 
\bar{q}_{1244} & \bar{q}_{3344} & 0 & 0 & 0 & 0 & 0 & 0\\
\hline
(2,2) & \bar{q}_{1122} & 0 & 0 & 0 & 0 & \bar{q}_{2244} & \bar{q}_{2222} & \bar{q}_{1222} & \bar{q}_{1222} & \bar{q}_{3322} & 0 & 0 & 0 & 0 & 0 & 0\\
\hline
(1,2) & 0 & 0 & 0 & 0 & 0 & \bar{q}_{1244} & \bar{q}_{1222} & \bar{q}_{1212} & \bar{q}_{1212} & \bar{q}_{1233} & 0 & 0 & 0 & 0 & 0 & 0\\
(2,1) & 0 & 0 & 0 & 0 & 0 & \bar{q}_{1244} & \bar{q}_{1222} & \bar{q}_{1212} & \bar{q}_{1212} & \bar{q}_{1233} & 0 & 0 & 0 & 0 & 0 & 0\\
\hline
(3,3) & \bar{q}_{1133} & 0 & 0 & 0 & 0 & \bar{q}_{3344} & \bar{q}_{3322} & \bar{q}_{1233}  & \bar{q}_{1233} & \bar{q}_{3333} & \bar{q}_{1333} & \bar{q}_{1333} & \bar{q}_{3233} & \bar{q}_{3233} & 0 & 0\\
\hline
(1,3) & 0 &  0 & 0 & 0 & 0 & 0 & 0 & 0 & 0 & \bar{q}_{1333} & \bar{q}_{1313} & \bar{q}_{1313} & \bar{q}_{1323} & \bar{q}_{1323} & 0 & 0\\
(3,1) & 0 & 0 & 0 & 0 & 0 & 0 & 0 & 0 & 0 & \bar{q}_{1333} & \bar{q}_{1313} & \bar{q}_{1313} & \bar{q}_{1323} & \bar{q}_{1323} & 0 & 0\\
(2,3) & 0 & 0 & 0 & 0 & 0 & 0 & 0 & 0 & 0 & \bar{q}_{3233} & \bar{q}_{1323} & \bar{q}_{1323} & \bar{q}_{3232} & \bar{q}_{3232} & 0 & 0\\
(3,2) & 0 & 0 & 0 & 0 & 0 & 0 & 0 & 0 & 0 & \bar{q}_{3233} & \bar{q}_{1323} & \bar{q}_{1323} & \bar{q}_{3232} & \bar{q}_{3232} & 0 & \\
\hline
(3,4) & 0 & 0 & 0 & 0 & 0 & 0 & 0 & 0 & 0 & 0 & 0 & 0 & 0 & 0 & 0 & 0\\
(4,3) & 0 & 0 & 0 & 0 & 0 & 0 & 0 & 0 & 0 & 0 & 0 & 0 & 0 & 0 & 0 & 0
\end{array}$$
\caption{\label{tab:flatt_q}Flattening matrix of $\bar{q}$ for a quartet $T=\{12\mid 34\}$. }
\label{table:flatq}
\end{table}
\end{landscape}

\begin{landscape}
\begin{table}
$$\begin{NiceArray}{c|c|cccc|c|c|cc|c|cccc|cc}
\CodeBefore
\rectanglecolor{gray!30}{3-3}{7-6} 
\rectanglecolor{gray!30}{7-9}{11-10} 
\rectanglecolor{gray!30}{11-12}{15-15} 
\rectanglecolor{gray!80}{3-3}{3-3} 
\rectanglecolor{gray!80}{9-9}{9-9}
\rectanglecolor{gray!80}{12-12}{12-12}
\Body
& (1,1) & (1,4) & (4,1) & (2,4) & (4,2) & (4,4) & (2,2) & (1,2) & (2,1) & (3,3) & (1,3) & (3,1) & (2,3) & (3,2) & (3,4) & (4,3)\\
\hline
(1,1) & \bar{p}_{1111} & 0 & 0 & 0 & 0 & \bar{p}_{1144} & \bar{p}_{1122} & 0 & 0 & \bar{p}_{1133} & 0 & 0 & 0 & 0 & 0 & 0\\
\hline
(1,4) & 0 & \bar{p}_{1414} & \bar{p}_{1441} & \bar{p}_{1424} & \bar{p}_{1442} & \bar{p}_{1444} & 0 & 0 & 0 & 0 & 0 & 0 & 0 & 0 & 0 & 0\\
(4,1) & 0 & \bar{p}_{4114} & \bar{p}_{4141} & \bar{p}_{4124} & \bar{p}_{4142} & \bar{p}_{4144} & 0 & 0 & 0 & 0 & 0 & 0 & 0 & 0 & 0 & 0\\
(2,4) & 0 & \bar{p}_{2414} & \bar{p}_{2441} & \bar{p}_{2424} & \bar{p}_{2442} & \bar{p}_{2444} & 0 & 0 & 0 & 0 & 0 & 0 & 0 & 0 & 0 & 0\\
(4,2) & 0 & \bar{p}_{4214} & \bar{p}_{4241} & \bar{p}_{4224} & \bar{p}_{4242} & \bar{p}_{4244} & 0 & 0 & 0 & 0 & 0 & 0 & 0 & 0 & 0 & 0\\
\hline
(4,4) & \bar{p}_{4411} & \bar{p}_{4414} & \bar{p}_{4441} & \bar{p}_{4424} & \bar{p}_{4442} & \bar{p}_{4444} & \bar{p}_{4422} & \bar{p}_{4412} & 
\bar{p}_{4421} & \bar{p}_{4433} & 0 & 0 & 0 & 0 & 0 & 0\\
\hline
(2,2) & \bar{p}_{2211} & 0 & 0 & 0 & 0 & \bar{p}_{2244} & \bar{p}_{2222} & \bar{p}_{2212} & \bar{p}_{2221} & \bar{p}_{2233} & 0 & 0 & 0 & 0 & 0 & 0\\
\hline
(1,2) & 0 & 0 & 0 & 0 & 0 & \bar{p}_{1244} & \bar{p}_{1222} & \bar{p}_{1212} & \bar{p}_{1221} & \bar{p}_{1233} & 0 & 0 & 0 & 0 & 0 & 0\\
(2,1) & 0 & 0 & 0 & 0 & 0 & \bar{p}_{2144} & \bar{p}_{2122} & \bar{p}_{2112} & \bar{p}_{2121} & \bar{p}_{2133} & 0 & 0 & 0 & 0 & 0 & 0\\
\hline
(3,3) & \bar{p}_{3311} & 0 & 0 & 0 & 0 & \bar{p}_{3344} & \bar{p}_{3322} & \bar{p}_{3312}  & \bar{p}_{3321} & \bar{p}_{3333} & \bar{p}_{3313} & \bar{p}_{3331} & \bar{p}_{3323} & \bar{p}_{3332} & 0 & 0\\
\hline
(1,3) & 0 &  0 & 0 & 0 & 0 & 0 & 0 & 0 & 0 & \bar{p}_{1333} & \bar{p}_{1313} & \bar{p}_{1331} & \bar{p}_{1323} & \bar{p}_{1332} & 0 & 0\\
(3,1) & 0 & 0 & 0 & 0 & 0 & 0 & 0 & 0 & 0 & \bar{p}_{3133} & \bar{p}_{3113} & \bar{p}_{3131} & \bar{p}_{3123} & \bar{p}_{3132} & 0 & 0\\
(2,3) & 0 & 0 & 0 & 0 & 0 & 0 & 0 & 0 & 0 & \bar{p}_{2333} & \bar{p}_{2313} & \bar{p}_{2331} & \bar{p}_{2323} & \bar{p}_{2332} & 0 & 0\\
(3,2) & 0 & 0 & 0 & 0 & 0 & 0 & 0 & 0 & 0 & \bar{p}_{3233} & \bar{p}_{3213} & \bar{p}_{3231} & \bar{p}_{3223} & \bar{p}_{3232} & 0 & \\
\hline
(3,4) & 0 & 0 & 0 & 0 & 0 & 0 & 0 & 0 & 0 & 0 & 0 & 0 & 0 & 0 & 0 & 0\\
(4,3) & 0 & 0 & 0 & 0 & 0 & 0 & 0 & 0 & 0 & 0 & 0 & 0 & 0 & 0 & 0 & 0
\end{NiceArray}$$
\caption{Flattening matrix of $\Bar{p}$ for a quartet $T=\{12\mid 34\}$. The highlighted submatrices have rank 1 as stated in~\Cref{lem:rank}. The entries highlighted in dark gray are generically non-zero. Therefore, the 18 quadratic edge invariants in \Cref{lem:localCIquartet} arise from considering all 2-minors of the grey submatrices containing the dark grey entry.}
\label{tab:quadrics}
\end{table}

\begin{table}
$$\begin{NiceArray}{c|c|cccc|c|c|cc|c|cccc|cc}
\CodeBefore
\rectanglecolor{gray!30}{2-2}{2-2} 
\rectanglecolor{gray!30}{7-2}{11-2} 
\rectanglecolor{gray!30}{2-8}{2-9} 
\rectanglecolor{gray!30}{7-8}{11-9} 
\rectanglecolor{gray!80}{2-2}{2-2} 
\rectanglecolor{gray!80}{2-9}{2-9}
\rectanglecolor{gray!80}{9-2}{9-2} 
\rectanglecolor{gray!80}{9-9}{9-9}
\Body
& (1,1) & (1,4) & (4,1) & (2,4) & (4,2) & (4,4) & (2,2) & (1,2) & (2,1) & (3,3) & (1,3) & (3,1) & (2,3) & (3,2) & (3,4) & (4,3)\\
\hline
(1,1) & \bar{p}_{1111} & 0 & 0 & 0 & 0 & \bar{p}_{1144} & \bar{p}_{1122} & 0 & 0 & \bar{p}_{1133} & 0 & 0 & 0 & 0 & 0 & 0\\
\hline
(1,4) & 0 & \bar{p}_{1414} & \bar{p}_{1441} & \bar{p}_{1424} & \bar{p}_{1442} & \bar{p}_{1444} & 0 & 0 & 0 & 0 & 0 & 0 & 0 & 0 & 0 & 0\\
(4,1) & 0 & \bar{p}_{4114} & \bar{p}_{4141} & \bar{p}_{4124} & \bar{p}_{4142} & \bar{p}_{4144} & 0 & 0 & 0 & 0 & 0 & 0 & 0 & 0 & 0 & 0\\
(2,4) & 0 & \bar{p}_{2414} & \bar{p}_{2441} & \bar{p}_{2424} & \bar{p}_{2442} & \bar{p}_{2444} & 0 & 0 & 0 & 0 & 0 & 0 & 0 & 0 & 0 & 0\\
(4,2) & 0 & \bar{p}_{4214} & \bar{p}_{4241} & \bar{p}_{4224} & \bar{p}_{4242} & \bar{p}_{4244} & 0 & 0 & 0 & 0 & 0 & 0 & 0 & 0 & 0 & 0\\
\hline
(4,4) & \bar{p}_{4411} & \bar{p}_{4414} & \bar{p}_{4441} & \bar{p}_{4424} & \bar{p}_{4442} & \bar{p}_{4444} & \bar{p}_{4422} & \bar{p}_{4412} & 
\bar{p}_{4421} & \bar{p}_{4433} & 0 & 0 & 0 & 0 & 0 & 0\\
\hline
(2,2) & \bar{p}_{2211} & 0 & 0 & 0 & 0 & \bar{p}_{2244} & \bar{p}_{2222} & \bar{p}_{2212} & \bar{p}_{2221} & \bar{p}_{2233} & 0 & 0 & 0 & 0 & 0 & 0\\
\hline
(1,2) & 0 & 0 & 0 & 0 & 0 & \bar{p}_{1244} & \bar{p}_{1222} & \bar{p}_{1212} & \bar{p}_{1221} & \bar{p}_{1233} & 0 & 0 & 0 & 0 & 0 & 0\\
(2,1) & 0 & 0 & 0 & 0 & 0 & \bar{p}_{2144} & \bar{p}_{2122} & \bar{p}_{2112} & \bar{p}_{2121} & \bar{p}_{2133} & 0 & 0 & 0 & 0 & 0 & 0\\
\hline
(3,3) & \bar{p}_{3311} & 0 & 0 & 0 & 0 & \bar{p}_{3344} & \bar{p}_{3322} & \bar{p}_{3312}  & \bar{p}_{3321} & \bar{p}_{3333} & \bar{p}_{3313} & \bar{p}_{3331} & \bar{p}_{3323} & \bar{p}_{3332} & 0 & 0\\
\hline
(1,3) & 0 &  0 & 0 & 0 & 0 & 0 & 0 & 0 & 0 & \bar{p}_{1333} & \bar{p}_{1313} & \bar{p}_{1331} & \bar{p}_{1323} & \bar{p}_{1332} & 0 & 0\\
(3,1) & 0 & 0 & 0 & 0 & 0 & 0 & 0 & 0 & 0 & \bar{p}_{3133} & \bar{p}_{3113} & \bar{p}_{3131} & \bar{p}_{3123} & \bar{p}_{3132} & 0 & 0\\
(2,3) & 0 & 0 & 0 & 0 & 0 & 0 & 0 & 0 & 0 & \bar{p}_{2333} & \bar{p}_{2313} & \bar{p}_{2331} & \bar{p}_{2323} & \bar{p}_{2332} & 0 & 0\\
(3,2) & 0 & 0 & 0 & 0 & 0 & 0 & 0 & 0 & 0 & \bar{p}_{3233} & \bar{p}_{3213} & \bar{p}_{3231} & \bar{p}_{3223} & \bar{p}_{3232} & 0 & \\
\hline
(3,4) & 0 & 0 & 0 & 0 & 0 & 0 & 0 & 0 & 0 & 0 & 0 & 0 & 0 & 0 & 0 & 0\\
(4,3) & 0 & 0 & 0 & 0 & 0 & 0 & 0 & 0 & 0 & 0 & 0 & 0 & 0 & 0 & 0 & 0
\end{NiceArray}$$
\caption{Flattening matrix of $\Bar{p}$ for a quartet $T=12\mid 34$. 
The entries highlighted in dark gray form a (generically) non-vanishing 2-minor and the submatrix formed by the entries highlighted in (dark and light) gray has rank 2. Therefore, the 4 cubic edge invariants in \Cref{lem:localCIquartet} arise from considering all 3-minors of the grey submatrix containing the dark grey minor.}
\label{tab:cubics}
\end{table}

\begin{table}
$$\begin{NiceArray}{c|c|cccc|c|c|cc|c|cccc|cc}
\CodeBefore
\rectanglecolor{gray!30}{2-2}{2-2} 
\rectanglecolor{gray!30}{7-2}{15-2} 
\rectanglecolor{gray!30}{2-9}{2-9} 
\rectanglecolor{gray!30}{7-9}{15-9} 
\rectanglecolor{gray!30}{2-11}{2-12} 
\rectanglecolor{gray!30}{7-11}{15-12} 
\rectanglecolor{gray!80}{2-2}{2-2} 
\rectanglecolor{gray!80}{2-9}{2-9}
\rectanglecolor{gray!80}{2-12}{2-12}
\rectanglecolor{gray!80}{9-2}{9-2} 
\rectanglecolor{gray!80}{12-2}{12-2}
\rectanglecolor{gray!80}{9-9}{9-9} 
\rectanglecolor{gray!80}{12-9}{12-9}
\rectanglecolor{gray!80}{9-12}{9-12} 
\rectanglecolor{gray!80}{12-12}{12-12}
\Body
& (1,1) & (1,4) & (4,1) & (2,4) & (4,2) & (4,4) & (2,2) & (1,2) & (2,1) & (3,3) & (1,3) & (3,1) & (2,3) & (3,2) & (3,4) & (4,3)\\
\hline
(1,1) & \bar{p}_{1111} & 0 & 0 & 0 & 0 & \bar{p}_{1144} & \bar{p}_{1122} & 0 & 0 & \bar{p}_{1133} & 0 & 0 & 0 & 0 & 0 & 0\\
\hline
(1,4) & 0 & \bar{p}_{1414} & \bar{p}_{1441} & \bar{p}_{1424} & \bar{p}_{1442} & \bar{p}_{1444} & 0 & 0 & 0 & 0 & 0 & 0 & 0 & 0 & 0 & 0\\
(4,1) & 0 & \bar{p}_{4114} & \bar{p}_{4141} & \bar{p}_{4124} & \bar{p}_{4142} & \bar{p}_{4144} & 0 & 0 & 0 & 0 & 0 & 0 & 0 & 0 & 0 & 0\\
(2,4) & 0 & \bar{p}_{2414} & \bar{p}_{2441} & \bar{p}_{2424} & \bar{p}_{2442} & \bar{p}_{2444} & 0 & 0 & 0 & 0 & 0 & 0 & 0 & 0 & 0 & 0\\
(4,2) & 0 & \bar{p}_{4214} & \bar{p}_{4241} & \bar{p}_{4224} & \bar{p}_{4242} & \bar{p}_{4244} & 0 & 0 & 0 & 0 & 0 & 0 & 0 & 0 & 0 & 0\\
\hline
(4,4) & \bar{p}_{4411} & \bar{p}_{4414} & \bar{p}_{4441} & \bar{p}_{4424} & \bar{p}_{4442} & \bar{p}_{4444} & \bar{p}_{4422} & \bar{p}_{4412} & 
\bar{p}_{4421} & \bar{p}_{4433} & 0 & 0 & 0 & 0 & 0 & 0\\
\hline
(2,2) & \bar{p}_{2211} & 0 & 0 & 0 & 0 & \bar{p}_{2244} & \bar{p}_{2222} & \bar{p}_{2212} & \bar{p}_{2221} & \bar{p}_{2233} & 0 & 0 & 0 & 0 & 0 & 0\\
\hline
(1,2) & 0 & 0 & 0 & 0 & 0 & \bar{p}_{1244} & \bar{p}_{1222} & \bar{p}_{1212} & \bar{p}_{1221} & \bar{p}_{1233} & 0 & 0 & 0 & 0 & 0 & 0\\
(2,1) & 0 & 0 & 0 & 0 & 0 & \bar{p}_{2144} & \bar{p}_{2122} & \bar{p}_{2112} & \bar{p}_{2121} & \bar{p}_{2133} & 0 & 0 & 0 & 0 & 0 & 0\\
\hline
(3,3) & \bar{p}_{3311} & 0 & 0 & 0 & 0 & \bar{p}_{3344} & \bar{p}_{3322} & \bar{p}_{3312}  & \bar{p}_{3321} & \bar{p}_{3333} & \bar{p}_{3313} & \bar{p}_{3331} & \bar{p}_{3323} & \bar{p}_{3332} & 0 & 0\\
\hline
(1,3) & 0 &  0 & 0 & 0 & 0 & 0 & 0 & 0 & 0 & \bar{p}_{1333} & \bar{p}_{1313} & \bar{p}_{1331} & \bar{p}_{1323} & \bar{p}_{1332} & 0 & 0\\
(3,1) & 0 & 0 & 0 & 0 & 0 & 0 & 0 & 0 & 0 & \bar{p}_{3133} & \bar{p}_{3113} & \bar{p}_{3131} & \bar{p}_{3123} & \bar{p}_{3132} & 0 & 0\\
(2,3) & 0 & 0 & 0 & 0 & 0 & 0 & 0 & 0 & 0 & \bar{p}_{2333} & \bar{p}_{2313} & \bar{p}_{2331} & \bar{p}_{2323} & \bar{p}_{2332} & 0 & 0\\
(3,2) & 0 & 0 & 0 & 0 & 0 & 0 & 0 & 0 & 0 & \bar{p}_{3233} & \bar{p}_{3213} & \bar{p}_{3231} & \bar{p}_{3223} & \bar{p}_{3232} & 0 & \\
\hline
(3,4) & 0 & 0 & 0 & 0 & 0 & 0 & 0 & 0 & 0 & 0 & 0 & 0 & 0 & 0 & 0 & 0\\
(4,3) & 0 & 0 & 0 & 0 & 0 & 0 & 0 & 0 & 0 & 0 & 0 & 0 & 0 & 0 & 0 & 0
\end{NiceArray}$$
\caption{Flattening matrix of $\Bar{p}$ for a quartet $T=12\mid 34$. 
The entries highlighted in dark gray form a (generically) non-vanishing 3-minor and the submatrix formed by the entries highlighted in (dark and light) gray has rank 3. Therefore, 7 of the quartic edge invariants in \Cref{lem:localCIquartet} arise from considering all 4-minors of the grey submatrix containing the dark grey minor.}
\label{tab:quarticsC}
\end{table}

\begin{table}
$$\begin{NiceArray}{c|c|cccc|c|c|cc|c|cccc|cc}
\CodeBefore
\rectanglecolor{gray!30}{2-2}{11-3} 
\rectanglecolor{gray!30}{2-9}{11-9} 
\rectanglecolor{gray!30}{2-7}{11-7} 
\rectanglecolor{gray!80}{2-2}{3-3} 
\rectanglecolor{gray!80}{2-9}{3-9}
\rectanglecolor{gray!80}{9-2}{9-3}
\rectanglecolor{gray!80}{9-9}{9-9}
\Body
& (1,1) & (1,4) & (4,1) & (2,4) & (4,2) & (4,4) & (2,2) & (1,2) & (2,1) & (3,3) & (1,3) & (3,1) & (2,3) & (3,2) & (3,4) & (4,3)\\
\hline
(1,1) & \bar{p}_{1111} & 0 & 0 & 0 & 0 & \bar{p}_{1144} & \bar{p}_{1122} & 0 & 0 & \bar{p}_{1133} & 0 & 0 & 0 & 0 & 0 & 0\\
\hline
(1,4) & 0 & \bar{p}_{1414} & \bar{p}_{1441} & \bar{p}_{1424} & \bar{p}_{1442} & \bar{p}_{1444} & 0 & 0 & 0 & 0 & 0 & 0 & 0 & 0 & 0 & 0\\
(4,1) & 0 & \bar{p}_{4114} & \bar{p}_{4141} & \bar{p}_{4124} & \bar{p}_{4142} & \bar{p}_{4144} & 0 & 0 & 0 & 0 & 0 & 0 & 0 & 0 & 0 & 0\\
(2,4) & 0 & \bar{p}_{2414} & \bar{p}_{2441} & \bar{p}_{2424} & \bar{p}_{2442} & \bar{p}_{2444} & 0 & 0 & 0 & 0 & 0 & 0 & 0 & 0 & 0 & 0\\
(4,2) & 0 & \bar{p}_{4214} & \bar{p}_{4241} & \bar{p}_{4224} & \bar{p}_{4242} & \bar{p}_{4244} & 0 & 0 & 0 & 0 & 0 & 0 & 0 & 0 & 0 & 0\\
\hline
(4,4) & \bar{p}_{4411} & \bar{p}_{4414} & \bar{p}_{4441} & \bar{p}_{4424} & \bar{p}_{4442} & \bar{p}_{4444} & \bar{p}_{4422} & \bar{p}_{4412} & 
\bar{p}_{4421} & \bar{p}_{4433} & 0 & 0 & 0 & 0 & 0 & 0\\
\hline
(2,2) & \bar{p}_{2211} & 0 & 0 & 0 & 0 & \bar{p}_{2244} & \bar{p}_{2222} & \bar{p}_{2212} & \bar{p}_{2221} & \bar{p}_{2233} & 0 & 0 & 0 & 0 & 0 & 0\\
\hline
(1,2) & 0 & 0 & 0 & 0 & 0 & \bar{p}_{1244} & \bar{p}_{1222} & \bar{p}_{1212} & \bar{p}_{1221} & \bar{p}_{1233} & 0 & 0 & 0 & 0 & 0 & 0\\
(2,1) & 0 & 0 & 0 & 0 & 0 & \bar{p}_{2144} & \bar{p}_{2122} & \bar{p}_{2112} & \bar{p}_{2121} & \bar{p}_{2133} & 0 & 0 & 0 & 0 & 0 & 0\\
\hline
(3,3) & \bar{p}_{3311} & 0 & 0 & 0 & 0 & \bar{p}_{3344} & \bar{p}_{3322} & \bar{p}_{3312}  & \bar{p}_{3321} & \bar{p}_{3333} & \bar{p}_{3313} & \bar{p}_{3331} & \bar{p}_{3323} & \bar{p}_{3332} & 0 & 0\\
\hline
(1,3) & 0 &  0 & 0 & 0 & 0 & 0 & 0 & 0 & 0 & \bar{p}_{1333} & \bar{p}_{1313} & \bar{p}_{1331} & \bar{p}_{1323} & \bar{p}_{1332} & 0 & 0\\
(3,1) & 0 & 0 & 0 & 0 & 0 & 0 & 0 & 0 & 0 & \bar{p}_{3133} & \bar{p}_{3113} & \bar{p}_{3131} & \bar{p}_{3123} & \bar{p}_{3132} & 0 & 0\\
(2,3) & 0 & 0 & 0 & 0 & 0 & 0 & 0 & 0 & 0 & \bar{p}_{2333} & \bar{p}_{2313} & \bar{p}_{2331} & \bar{p}_{2323} & \bar{p}_{2332} & 0 & 0\\
(3,2) & 0 & 0 & 0 & 0 & 0 & 0 & 0 & 0 & 0 & \bar{p}_{3233} & \bar{p}_{3213} & \bar{p}_{3231} & \bar{p}_{3223} & \bar{p}_{3232} & 0 & \\
\hline
(3,4) & 0 & 0 & 0 & 0 & 0 & 0 & 0 & 0 & 0 & 0 & 0 & 0 & 0 & 0 & 0 & 0\\
(4,3) & 0 & 0 & 0 & 0 & 0 & 0 & 0 & 0 & 0 & 0 & 0 & 0 & 0 & 0 & 0 & 0
\end{NiceArray}$$
\caption{Flattening matrix of $\Bar{p}$ for a quartet $T=12\mid 34$. 
The entries highlighted in dark gray form a (generically) non-vanishing 3-minor and the submatrix formed by the entries highlighted in (dark and light) gray has rank 3. Therefore, 7 of the quartic edge invariants in \Cref{lem:localCIquartet} arise from considering all 4-minors of the grey submatrix containing the dark grey minor.}
\label{tab:quarticsT}
\end{table}
\end{landscape}

\end{document}